\documentclass[showpacs,aps,prd,eqsecnum,floats]{revtex4}

\usepackage[T1]{fontenc}
\usepackage{graphicx}
\usepackage{vmargin}

\input epsf                                                            
\usepackage{graphics}
\def\be{\begin{equation}}
\def\ee{\end{equation}}
\def\bea{\begin{eqnarray}}
  \def\eea{\end{eqnarray}}

\def\lsim{\mathrel{\rlap{\lower4pt\hbox{\hskip1pt$\sim$}}
    \raise1pt\hbox{$<$}}}                % less than or approx. symbol
\def\gsim{\mathrel{\rlap{\lower4pt\hbox{\hskip1pt$\sim$}}
    \raise1pt\hbox{$>$}}}                % greater than or approx. symbol

\begin{document}

\title{Primordial power spectrum from WMAP}

\author{Arman Shafieloo \email{arman@iucaa.ernet.in}}
\author{Tarun Souradeep \email{tarun@iucaa.ernet.in}}

\affiliation{Inter-University Centre for Astronomy and Astrophysics,
Post Bag 4, Ganeshkhind, Pune 411~007, India.}

\begin{abstract}

The observed angular power spectrum of the cosmic microwave background
temperature anisotropy, $C_l$, is a convolution of a cosmological
radiative transport kernel with an assumed primordial power spectrum
of inhomogeneities. Exquisite measurements of $C_l$ over a wide range
of multipoles from the Wilkinson Microwave Anisotropy Probe (WMAP) has
opened up the possibility to deconvolve the primordial power spectrum
for a given set of cosmological parameters (base model).  We implement
an improved (error sensitive) Richardson-Lucy deconvolution algorithm
on the measured angular power spectrum from WMAP assuming the
concordance cosmological model.  The most prominent feature of the
recovered $P(k)$ is a sharp, infra-red cut off on the horizon
scale. The resultant $C_l$ spectrum using the recovered spectrum has a
likelihood far better than a scale invariant, or, `best fit' scale
free spectra ($\Delta\ln{\cal L}\approx 25$ {\it w.r.t.} Harrison
Zeldovich, and, $\Delta\ln{\cal L}\approx 11$ {\it w.r.t.} power law
with $n_s=0.95$). The recovered $P(k)$ has a localized excess just
above the cut-off which leads to great improvement of likelihood over
the simple monotonic forms of model infra-red cut-off spectra
considered in the post WMAP literature.  The recovered $P(k)$, in
particular, the form of infra-red cut-off is robust to small changes
in the cosmological parameters.  We show that remarkably similar form
of infra-red cutoff is known to arise in very reasonable extensions
and refinements of the predictions from simple inflationary scenarios.
Our method can be extended to other cosmological observations such as
the measured matter power spectrum and, in particular, the much
awaited polarization spectrum from WMAP.

\end{abstract}
\maketitle

\section{Introduction}

Increasingly accurate measurements of the anisotropy in the
temperature of the cosmic microwave background (CMB) has ushered in an
era of precision cosmology. A golden decade of CMB anisotropy
measurements by numerous experiments was topped by the results from
the first year of data obtained by the Wilkinson Microwave Anisotropy
Probe (WMAP) ~\cite{ben_wmap03}.  Under simple hypotheses for the
spectrum of primordial perturbations, exquisite estimates of the
cosmological parameters have been obtained from the angular power
spectrum measurement by WMAP combined with other cosmological
observations~\cite{sper_wmap03}. Although the assumed, scale free
(with mild deviations), initial power spectra may be a generic
prediction of the simplest scenarios of generation of perturbations
during inflation, initial spectra with radical deviations are known to
arise from very reasonable extensions, or, refinements to the simplest
scenarios~\cite{oldbsi,star92,vilfor82}.  Consequently, cosmological
parameter estimation from the CMB anisotropy and the matter power
spectrum obtained from redshift surveys, weak gravitational lensing
and Ly-$\alpha$ absorption, depends sensitively on the dimensionality,
nature and freedom in the parameter space of initial
conditions~\cite{uscosmo97}.

The angular power spectrum, $C_l$, is a convolution of the initial
power spectrum $P(k)$ generated in the early universe with a radiative
transport kernel, $G(l,k)$, that is determined by the current values
of the cosmological parameters.  The remarkably precise observations
of the angular power spectrum $C_l$ by WMAP, and the concordance of
cosmological parameters measured from different cosmological
observations opens up the avenue to directly recover the initial power
spectrum of the density perturbation from the observations. The
Richardson-Lucy (RL) method deconvolution was shown to be a promising
and powerful method to measure the power spectrum of initial
perturbations from the CMB angular power spectrum~\cite{arman_msc}. In
this paper, we apply the method to the CMB anisotropy spectrum
measured by WMAP.  We have also devised and implemented an improvement
to the RL scheme, whereby the iterative deconvolution algorithm is
designed to converge and {\em match the measurements only within the
given error-bars.}

The direct numerical deconvolution has clear advantages over the other
prevalent approach of obtaining the most likely parameter values for a
parametric model primordial spectra~\footnote{ Estimate of the power
spectrum in $k$ space `bins' carried out with recent data is a
somewhat model independent~\cite{brid_eft03,pia_wan03,han03}. Direct
deconvolution with different method~\cite{mat_sas0203} has been
attempted~\cite{kog03}}. First, as emerges from our work, the direct
method can reveal features that are not anticipated by the parametric
model spectra, hence would be completely missed out in the latter
approach.  Second, in the absence of an accepted early universe
scenario (more narrowly, a favored model of inflation), it is
difficult to {\it a priori} set up and justify the chosen space of
initial conditions. The complex covariances between the cosmological
and the initial parameters are sensitive to the parameterization of
the space of initial spectra adopted.  Efforts along these lines are
further obscured by issues such as the applicability of the Occam's
razor to dissuade extension of the parameter space of initial
conditions.  Such deliberations have been recently framed in the more
quantitative language of Bayesian evidence to evaluate and select
between possible parameterizations~\cite{jaf03}.  However, this
approach cannot really point to a preferred parameterization. Whereas,
in the direct approach we can evade the issue of appropriate
parameterization of the initial power spectrum. For a given set of
cosmological parameters, our method obtains the primordial power
spectrum that `maximizes' the likelihood. Hence, the space of
parameters remains that of the (more widely accepted and agreed up on)
cosmological parameters alone.  In principle, it is possible to
explore the entire space of cosmological parameters along the lines
being done routinely, and instead of simply computing the likelihood
for a given model initial power spectrum, one obtains the initial
power spectrum that maximizes the likelihood at that point and assigns
that likelihood to that point in the space of cosmological
parameters~\footnote{It is possible that very unlikely cosmological
parameters get picked out due to suitably tailored initial power
spectrum.  In this case, one can employ appropriately strong priors
from other observation or beliefs to ensure that unlikely, unphysical,
ill-motivated are down weighted.}.  However, in this paper we limit
ourselves to obtaining the primordial spectrum for a few given sets of
`best fit' cosmological parameters and showing that recovered $P(k)$
is robust to small local variations in the cosmological parameters. We
defer the wide exploration of the cosmological parameter space to
future work.

  In section~\ref{meth}, we describe the problem and present the
deconvolution method, details of its implementation and tests of
recovering known primordial spectra from synthetic angular power
spectrum data. The recovered spectrum from WMAP data using our method
is described in section~\ref{results}. The recovered primordial
spectrum has statistically significant features that are robust to the
variations in the cosmological parameters.  In section~\ref{disc}, the
recovered spectrum is shown to be tantalizingly similar to some forms
of broken scale invariant spectra known to arise from reasonable
variations of the simplest inflationary scenario.

\section{Method}
\label{meth}

\subsection{Integral equation for CMB anisotropy}

In this subsection we recall the integral equation for the angular
power spectrum of CMB anisotropy and set up the inverse problem that
we solve using a deconvolution method.  The observed CMB anisotropy
${\Delta_T}({\bf n})$ is one realization of a random field on the
surface of a sphere and can be expressed in terms of the random
variates $a_{lm}$ that are the coefficients of a Spherical Harmonic
expansion given by

\be {\Delta_T}({\bf
n})\,=\,\sum_{l=2}^{\infty}\,\sum_{m=-l}^{m=l}\,a_{lm}\,Y_{lm}({\bf
n}) 
\ee 
where $Y_{lm}({\bf n})$ are the Spherical Harmonic functions.
For an underlying isotropic, Gaussian statistics, the angular power
spectrum, $C_l$ defined through \be \langle a_{lm} a^*_{l^\prime
m^\prime} \rangle=C_{l} \delta_{ll^\prime} \delta_{mm^\prime}\,, \ee
completely characterizes the CMB anisotropy.

In a flat universe the temperature fluctuation in the CMB photons at the
location ${\bf x}$ at the present conformal time $\eta_0$ propagating
in a direction ${\bf n}$ is 

\be \Delta ({\bf n}) \equiv \,\Delta (x_0, {\bf n} ,\eta_0)\,=\,\int
d^3k\,e^{i\,{\bf k\cdot{\bf x}}}\,\Delta (k,{\bf n},\eta_0).  
\ee 

For globally isotropic cosmology, the temperature fluctuation $\Delta
({\bf k},{\bf n},\eta_0)\equiv \Delta ({\bf k}\cdot{\bf n},\eta_0)$ 
can be expanded in terms of Legendre polynomials leading to 

\be 
\Delta_T (x,{\bf n} ,\eta_0)\,=\,\int
d^3k\,e^{i\,{\bf k\cdot{\bf x}}}\,
\sum_{l=0}^{\infty}(-i)^l \,(2l+1) \Delta_l(k,\eta_0)
P_l({\bf k\cdot n})\,.
\ee 

The angular power spectrum $C_l$ given by the coefficients of Legendre
expansion is then expressed as

\be C_l\,=\,(4\pi)^2\,\int
\,\frac{dk}{k}\,P(k)\,|{\Delta_{Tl}(k,\eta_0)}|^2
\label{clintg}
\ee where $P(k)$, the power spectrum of the {\em primordial (scalar)
metric} initial perturbation $\psi_{\rm prim}$, is given by
 
\be \langle \psi_{\rm prim}({\bf k})\,\psi_{\rm prim}^*({\bf
k^\prime})\rangle \,=\,\frac{P(k)}{k^3}\,\delta({\bf k}-{\bf
k^\prime}) \ee because the $k$ space modes are uncorrelated in a
homogeneous space.  The spectrum $P(k)$ represents the {\it r.m.s.}
power in the scalar metric perturbations per logarithmic interval
$dk/k$ at a wavenumber $k$. It is related to power spectrum of the
primordial modes of density perturbations, $\delta_k$, as $P(k)
\propto |\delta_k|^2/k$.  For the conventional scale free
parameterization, the spectral index $n_s$ is defined through
$|\delta_k|^2= A k^{n_s}$. {\em The power spectrum $P(k)$ is a
constant for the scale invariant Harrison-Zeldovich spectrum
(corresponds to $n_s=1$).}

The harmonics of the temperature fluctuations at the current epoch,
$\Delta_{Tl}(k,\eta_0)$, is obtained from the solution to the
Boltzmann equation for the CMB photon distribution.  In this paper we
use $\Delta_{Tl}(k,\eta_0)$ computed by the CMBfast
software~\cite{uros_zal96}.  Numerically, a suitably discretized space
of wave-numbers, $k_i$ is used where the following discrete version of
integral eq.~(\ref{clintg}) is applicable 

\bea C_l\,= \sum_i G(l,k_i )\,P(k_i) \nonumber\\ G(l,k_i) = 
\frac{\Delta k_i}{k_i}\,|{\Delta_{Tl}(k_i,\eta_0)}|^2\,.
\label{clsum}
\eea In the above equation, the {\em `target'} angular power spectrum,
$C_l\equiv C_l^D$, is the data given by observations, and the
radiative transport kernel, $G(l,k)$ is fixed by the cosmological
parameters of the {\em `base'} model. (The kernel $G(l,k)$ also
includes the effect of geometrical projection from the three
dimensional wavenumber, $k$, to the harmonic multipole, $l$ on the two
dimensional sphere.) For a given $G(l,k)$, obtaining the primordial
power spectrum, $P(k)$ from the measured $C_l$ is clearly a
deconvolution problem. An important feature of our problem is that
$C_l^D$, $G(l,k)$ and $P(k)$ are all positive definite.  However, to
get reliable results from the deconvolution, we require high signal to
noise measurements of $C_l^D$ over a large range of multipoles with
good resolution in multipole, preferably, from a single experiment to
avoid the uncertainties of relative calibration~\footnote{Binned
$C_l^D$ data that combined the heterogeneous CMB band power obtained
from different experimental data sets are not always as reliable due
to relative calibration uncertainties.  Application of our method to
binned data of \cite{pod01} did not give robust or convincing results
~\cite{arman_msc}. However, in principle nothing rules out using
binned data from heterogeneous data sets that have good
cross-calibration.}.

Ideally, it will be best to measure each $C_l^D$ independently.  In
practice, incomplete sky coverage and other effects limit the
resolution in multipole space.  All experiments provide band power
estimates, $\bar C_b({l_{\rm eff}}) = \sum_l W_l^{(b)}\,C_l$ which are
averaged linear combinations of the underlying $C_l$. Since
eq.~(\ref{clsum}) is linear, a similar equation holds for all the band
powers with a kernel, $\bar G(l_{\rm eff},k_i) = \sum_l
W^{(b)}_l\,G(l,k_i)$.  {\em For simplicity and brevity of notation, we
retain the notation $C_l^D$ and $G(l,k)$ for band power estimates with
$l$ denoting the bin center. We also implicitly assume a discrete
wavenumber $k$ instead of carrying the clumsy notation, $k_i$.}

As mentioned in more detail in section~\ref{wmapdat}, the CMB angular
power spectrum from WMAP ranges from the quadrupole, $l=2$ to
$900$. Moreover, the `full sky' coverage of WMAP implies good
resolution in multipole space. We use WMAP TT
(temperature-temperature) binned power spectrum as $C_l^D$ in
eq.~(\ref{clsum}). We use CMBfast software to compute the the $G(l,k)$
matrix for the post-WMAP `best fit' cosmological parameters. We
emphasize that although the $P(k)$ is recovered using binned data, the
significance (performance) of the recovered spectrum is evaluated
using the WMAP likelihood of entire unbinned $C_l$ properly accounting
for covariances. The likelihood is computed using the numerical code,
data and its error covariance provided by the WMAP collaboration with
the release of the first year data.

In this work we limit our attention to the angular spectrum of the
temperature anisotropy, $C_l^{TT}$.  Including the polarization of CMB
photons, equations similar to eq.~(\ref{clintg}) can be also written
for the three additional angular power spectra, $C_l^{TE}$, $C_l^{EE}$
and $C_l^{BB}$ involving their corresponding kernels. It will
certainly be interesting to include these when more complete
polarization data is available in the future.  At present, only the
$C_l^{TE}$ spectrum has been published by the WMAP team which is not
positive definite and hence, not ideally suited for our
method. However, once $C_l^{EE}$ data is made available, our method
can readily accommodate both $C_l^{TE}$ and $C_l^{EE}$ together since
combinations such as $C_l^{TT} \pm 2C_l^{TE}+C_l^{EE}$ are positive
definite.

\subsection{Deconvolution method}

The Richardson-Lucy (RL) algorithm was developed and is widely used in
the context of image reconstruction in
astronomy~\cite{lucy74,rich72}. However, the method has also been
successfully used in cosmology, to deproject the $3$-D correlation
function and power spectrum from the measured $2$-D angular
correlation and $2$-D power spectrum~\cite{baug_efs93,baug_efs94}.  We
employ an improved RL method to solve the inverse problem for $P(k)$
in eq.~(\ref{clsum}).  The advantage of RL method is that positivity
of the recovered $P(k)$ is automatically ensured, given $G(l,k)$ is
positive definite and $C_l$'s are positive.

The RL method is readily derived from elementary probability theory on
distributions~\cite{lucy74}. To make this connection, we consider
normalized quantities~\footnote{In what follows we assume that the
quantities are normalized and omit the overhead tilde in the
notation. The normalization to unity not required and it is possible
to use other normalizations, such as, $\sum (2 l+1) C_l = constant$.}

\be \,\sum_l\, \tilde C_l\,=\,1; \quad \,\sum_k\, \tilde P(k)\,=\,1;
\quad \,\sum_l\, \tilde G(l,k)\,=\,1\,.  \ee 
This allows us to view the functions, $P(k)$ and $C_l$ as one
dimensional distributions, and $G(l,k)$ as a conditional probability
distribution.  Further, for convenience (not necessity) of writing an
infinitesimal measure $dl$ we view $l$ to be continuous.  The
integrand in the integral eq.~(\ref{clintg}) suggests defining two
other probability distributions $Q(l,k)$ and $L(l,k)$, such that

\be
G(l,k)\,P(k)\,dldk\,=\,Q(l,k)\,dldk\,=\,L(k,l)\,C_l\,dldk
\label{probeq}
\ee
Dividing the both side of eq.~(\ref{probeq}) by $C_l\,dldk$ we 
obtain

\be
L(k,l)\,=\,\frac{P(k)}{C_l}\,G(l,k)\,.
\label{rl2}
\ee
The normalization conditions imply 
\be
P(k)\,=\,\int Q(l,k)\,dl\,=\,\int C_l\,L(k,l)dl \,,
\ee
which in the discrete $l$ space reads

\be
P(k)\,=\,\sum_l\,Q(l,k)\,=\,\sum_l\,L(k,l)\,C_l\,.
\label{rl1}
\ee

 The RL method iteratively solves the eqs.~(\ref{rl2}) and
(\ref{rl1}). Starting from an initial guess $P^{(0)}(k)$, $L$ is
obtained using eq.~(\ref{rl2}) as the first step. The second step is
to obtain a revised $P^{(1)}(k)$ using eq.~(\ref{rl1}). These two
steps are repeated iteratively with the $P^{(i)}(k)$ obtained after
iteration $i$ feeding into the iteration $i+1$.  In principle, the
final answer could depend on the initial guess but in practice, for a
large variety of problems, RL is known leads to the correct answer
even a crude estimation of the initial guess. In particular, for our
problem the RL rapidly converges to the same solution $P(k)$
independent of the initial guess $P^{(0)}(k)$. This is demonstrated in
Appendix~\ref{initguess}.

The iterative method can be neatly encoded into a simple recurrence
relation.  The power spectrum $P^{(i+1)}(k)$ recovered after iteration
$(i+1)$ is given by

\be
P^{(i+1)}(k)- P^{(i)}(k)\,=\,P^{(i)}(k)\,\sum_l\,G(l,k)\,\frac
{C^D_l\,-C_l^{(i)}}{C_l^{(i)}}
\label{RLstd}
\ee 
where $C^D_l$ is the measured data (target) and $C_l^{(i)}$ is the
angular power spectrum at $i^{\rm th}$ iteration obtained from
eq.~(\ref{clsum}) using the recovered power spectrum $P^{(i)}(k)$.
Eq.~(\ref{RLstd}) with eq.~(\ref{clsum}) for obtaining $C_l^{(i)}$
from $P^{(i)}(k)$ completely summarizes the standard RL method.

Due to noise and sample variance, the data $C_l^D$ is measured within
some non-zero error bars $\sigma_l$. The standard RL method does not
incorporate the error information at all.  Consequently, a well known
drawback of the standard RL method is that at large iterations the
method starts fitting features from the noise.  Modified forms of RL
that address this issue have been proposed (e.g., see damped RL method
in \cite{dampRL}). In our problem, this problem manifests itself as
very non-smooth deconvolved spectrum $P(k)$ that has poor likelihood
with the full WMAP spectrum data.  We devise a novel method to make
the RL method sensitive to the error $\sigma_l$ by modifying
eq.~(\ref{RLstd}) to

\be P^{(i+1)}(k)- P^{(i)}(k)\,=\,P^{(i)}(k)\,\sum_l\,G(l,k)\,\frac
{C_l\,-C_l^{(i)}}{C_l^{(i)}}\,\,\,\tanh^2\,
\left[\frac{(C^D_l\,-C_l^{(i)})^2}{{\sigma_l}^2}\right].
\label{RLerr}
 \ee The idea is to employ a `convergence' function to progressively
weigh down the contribution to the correction $P^{(i+1)}-P^{(i)}$ from
a multipole bin where $C_l^{(i)}$ is close $C_l^D$ within the error
bar $\sigma_l$. This innovation significantly improves the WMAP
likelihood of the deconvolved spectrum. For certain $G(l,k)$, the
improvement is so dramatic that using IRL becomes very crucial to
successful recovery of the spectrum (see~\ref{paramvar}). The final
results are not sensitive to the exact functional form of the
convergence function. The choice given in eq.~(\ref{RLerr}) works well
but is not unique in any sense.

At every iteration of the IRL scheme, we compute the $\chi^2$ of the
$C_l^{(i)}$ with respect to the binned data $C_l^D$. We have found
that the IRL iterations (as well as the RL) march almost monotonically
toward improved (smaller) $\chi^2$. We halt the iterations when the
$\chi^2$ does not change appreciably in subsequent iterations of IRL.

\begin{figure}[h]
\includegraphics[scale=0.45, angle=-90]{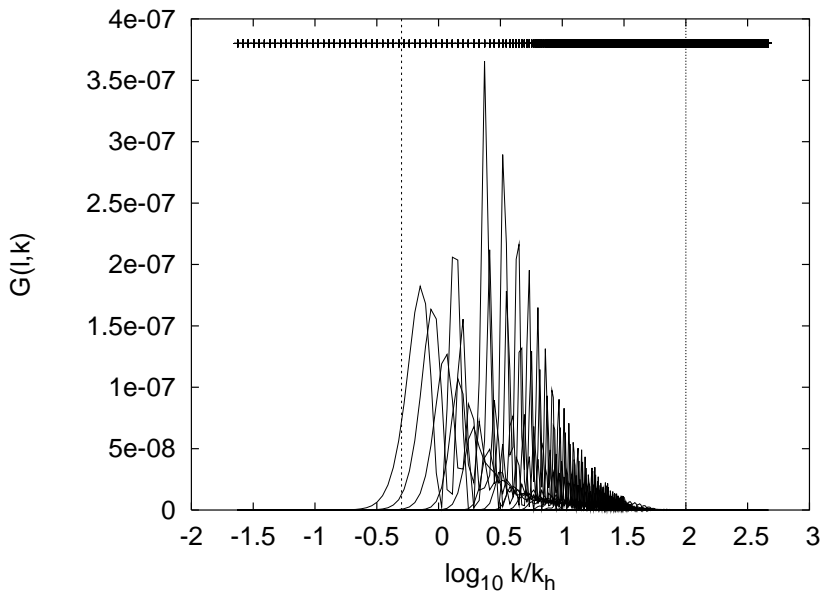}
  \caption{\small The curves are $\bar G(l,k)$ versus wavenumber $k$
used in our work.  $\bar G(l,k)$ is averaged over $G(l,k)$ within
multipole bins used by WMAP.  The two vertical lines roughly enclose
the region of $k$-space strongly probed by the kernel where the
primordial spectrum can be expected to be recovered reliably.  The
$k$-space sampling used is indicated by the line of `$+$' symbols at
the top of the plot. }
\label{glk}
\end{figure}

\subsection{Post processing the deconvolved spectrum }

The deconvolution algorithm produces a `raw' $P(k)$ that has to be
processed further.  The raw deconvolved spectrum has spurious
oscillations and features arising largely out of the $k$ space
sampling and binning in $l$ space. These numerical artifacts are
common to all recovered spectra recovered with same
$G(l,k)$. Fig.~\ref{glk} shows the plot of the binned kernel $G(l,k)$
for the $l$-space bins used by WMAP~\cite{hin_wmap03,expl_wmap03}. The
wavenumbers are scaled by horizon scale, $k_h = 2\pi/(\eta_0-\eta_{\rm
rec})$. The sampling of $k$ space used here is also indicated in the
figure. We find that removing these artifacts and smoothing the
resultant spectrum improves the WMAP likelihood of the corresponding
$C_l$.

\begin{figure*} 
\centering
\begin{center}
\vspace{-0.05in}
\centerline{\mbox{\hspace{0.in} \hspace{2.1in}  \hspace{2.1in} }}
$\begin{array}{@{\hspace{-0.4in}}c@{\hspace{0.3in}}c@{\hspace{0.3in}}c}
\multicolumn{1}{l}{\mbox{}} &
\multicolumn{1}{l}{\mbox{}} &
\multicolumn{1}{l}{\mbox{}} \\ [-0.5cm]
 
\includegraphics[scale=0.22, angle=-90]{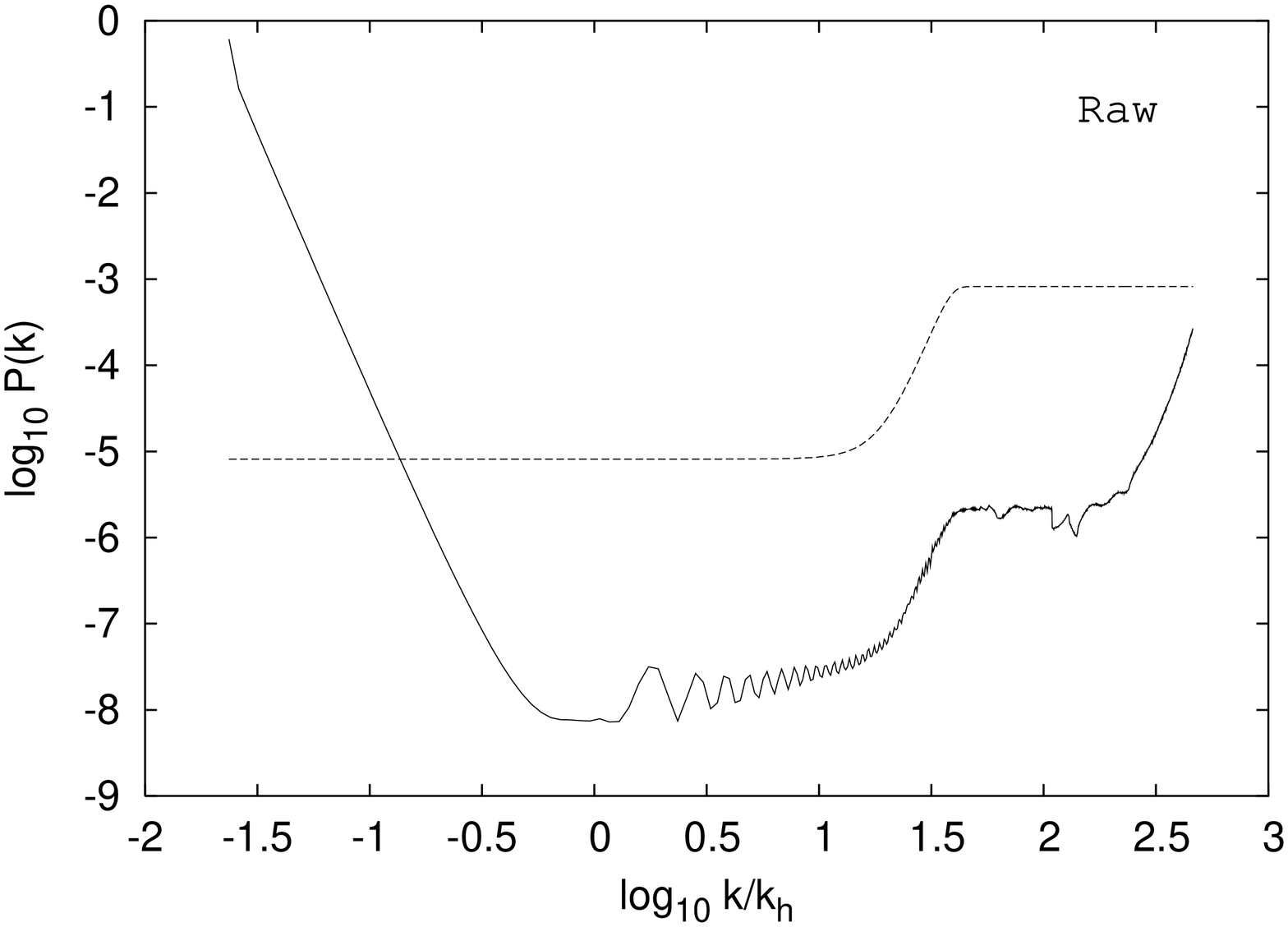}
 
\includegraphics[scale=0.22, angle=-90]{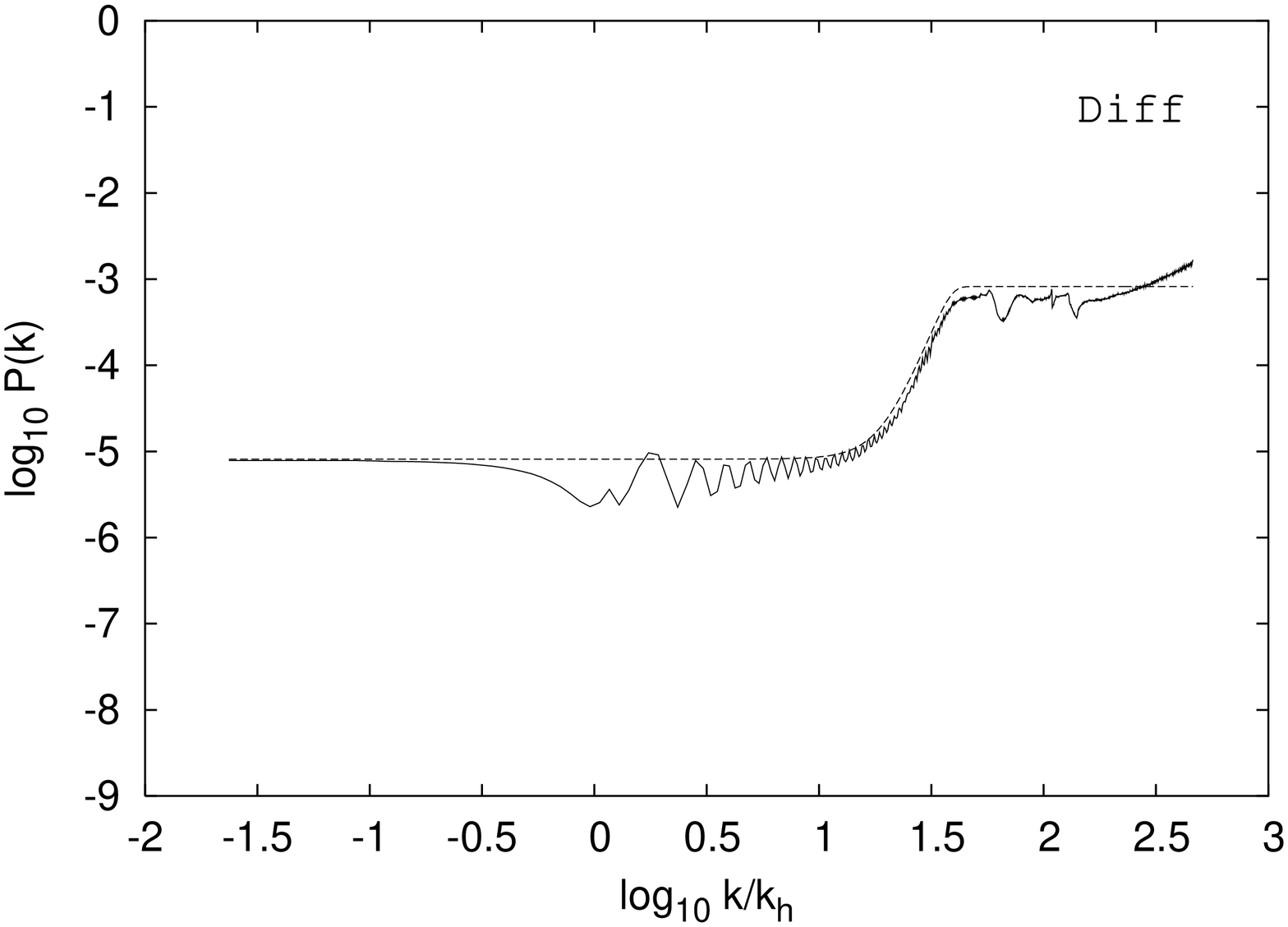}

\includegraphics[scale=0.22, angle=-90]{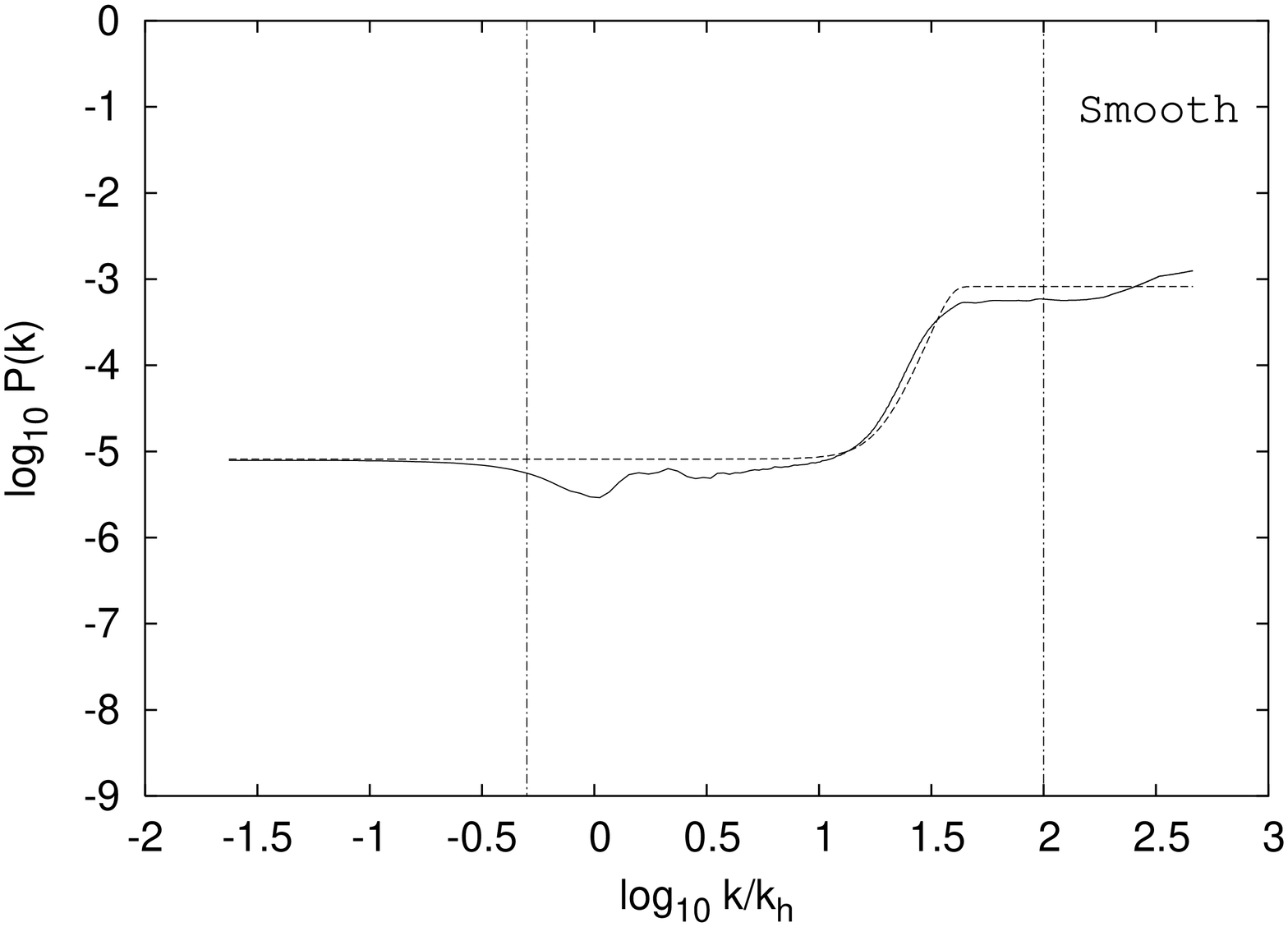}

%\mbox{\bf (a)} & \mbox{\bf (b)} & \mbox{\bf (c)}
\end{array}$

$\begin{array}{@{\hspace{-0.4in}}c@{\hspace{0.3in}}c@{\hspace{0.3in}}c}
\multicolumn{1}{l}{\mbox{}} &
\multicolumn{1}{l}{\mbox{}} &
\multicolumn{1}{l}{\mbox{}} \\ [-0.5cm]

\includegraphics[scale=0.22, angle=-90]{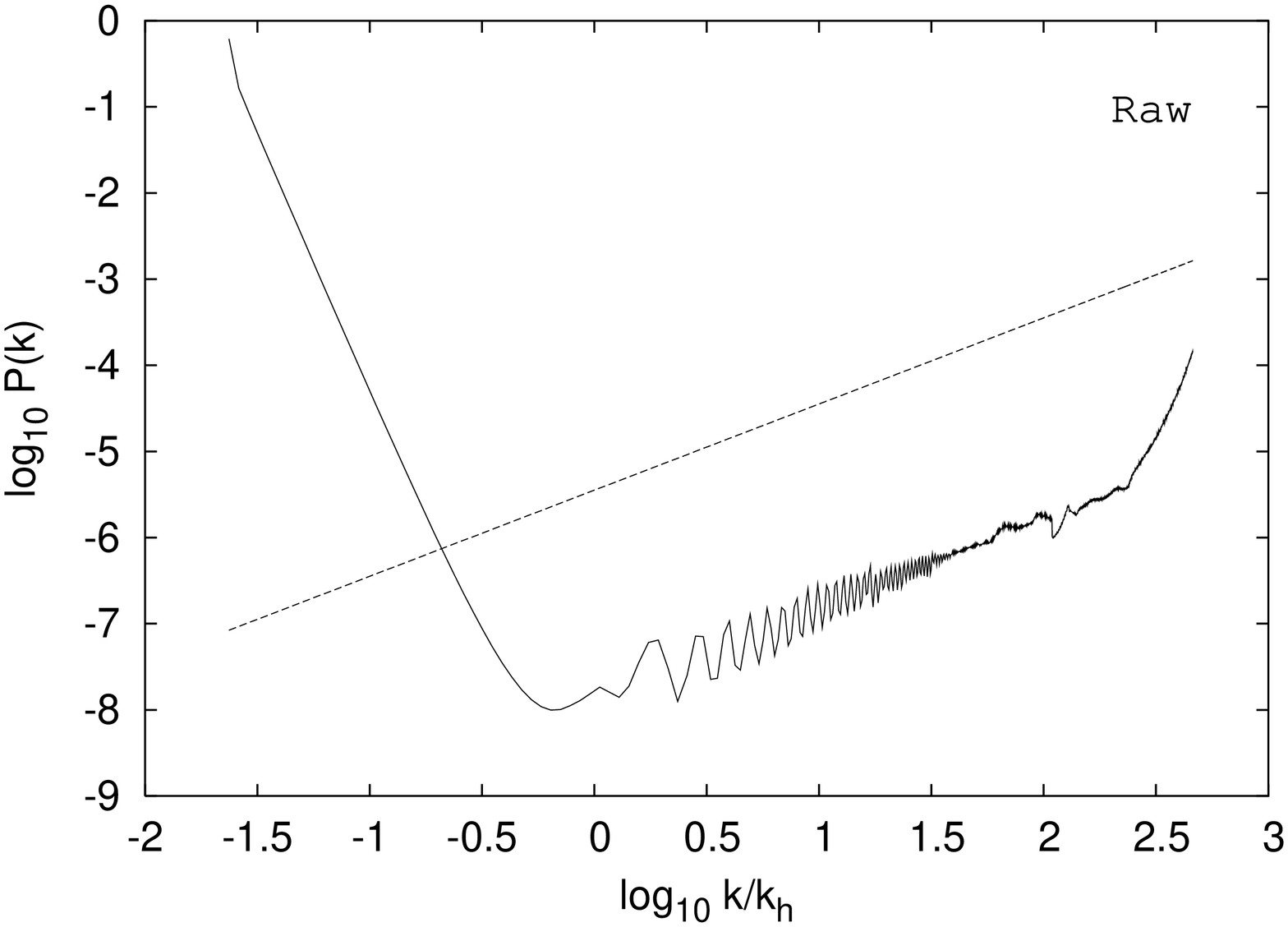}
  
\includegraphics[scale=0.22, angle=-90]{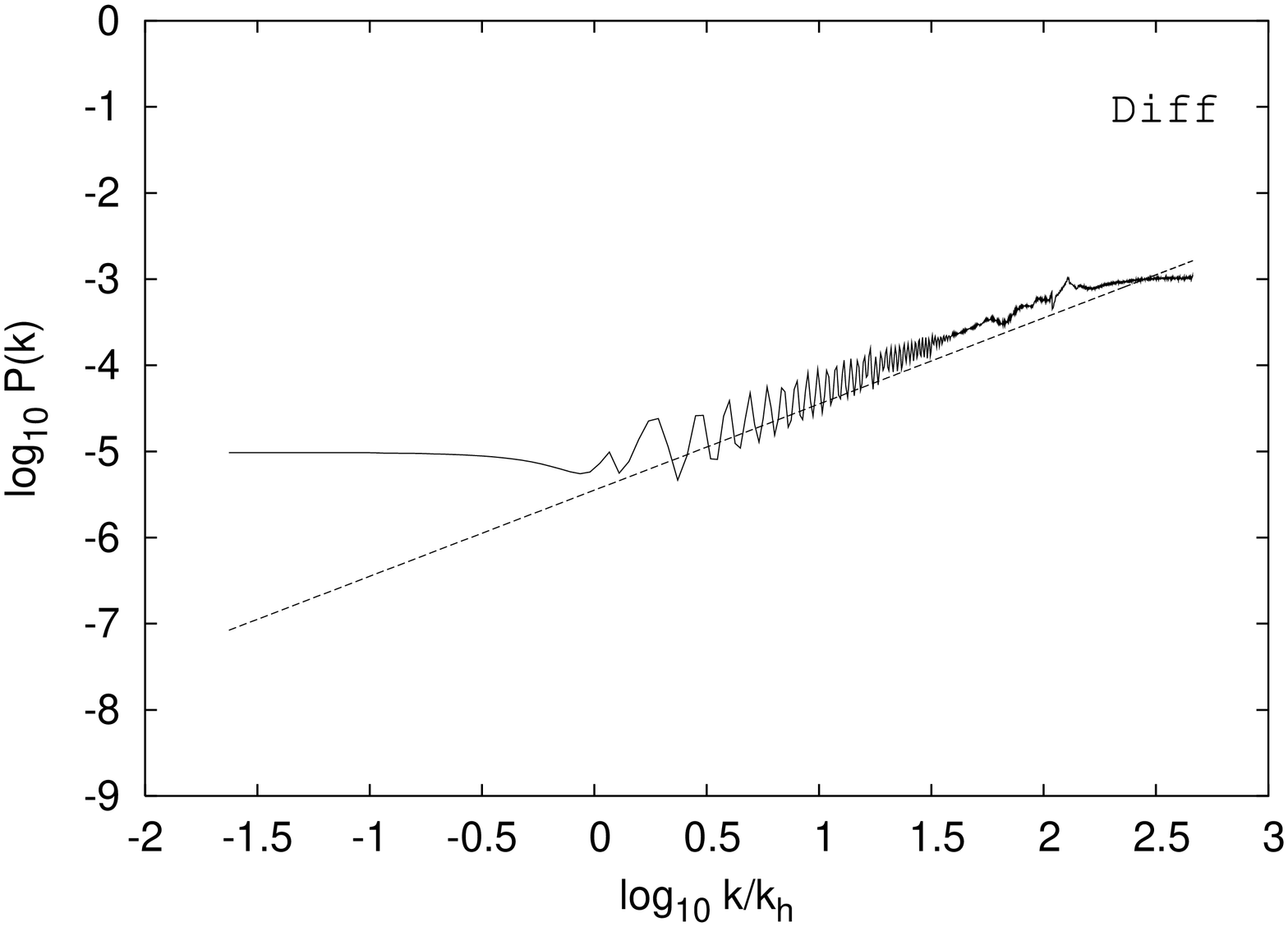}

\includegraphics[scale=0.22, angle=-90]{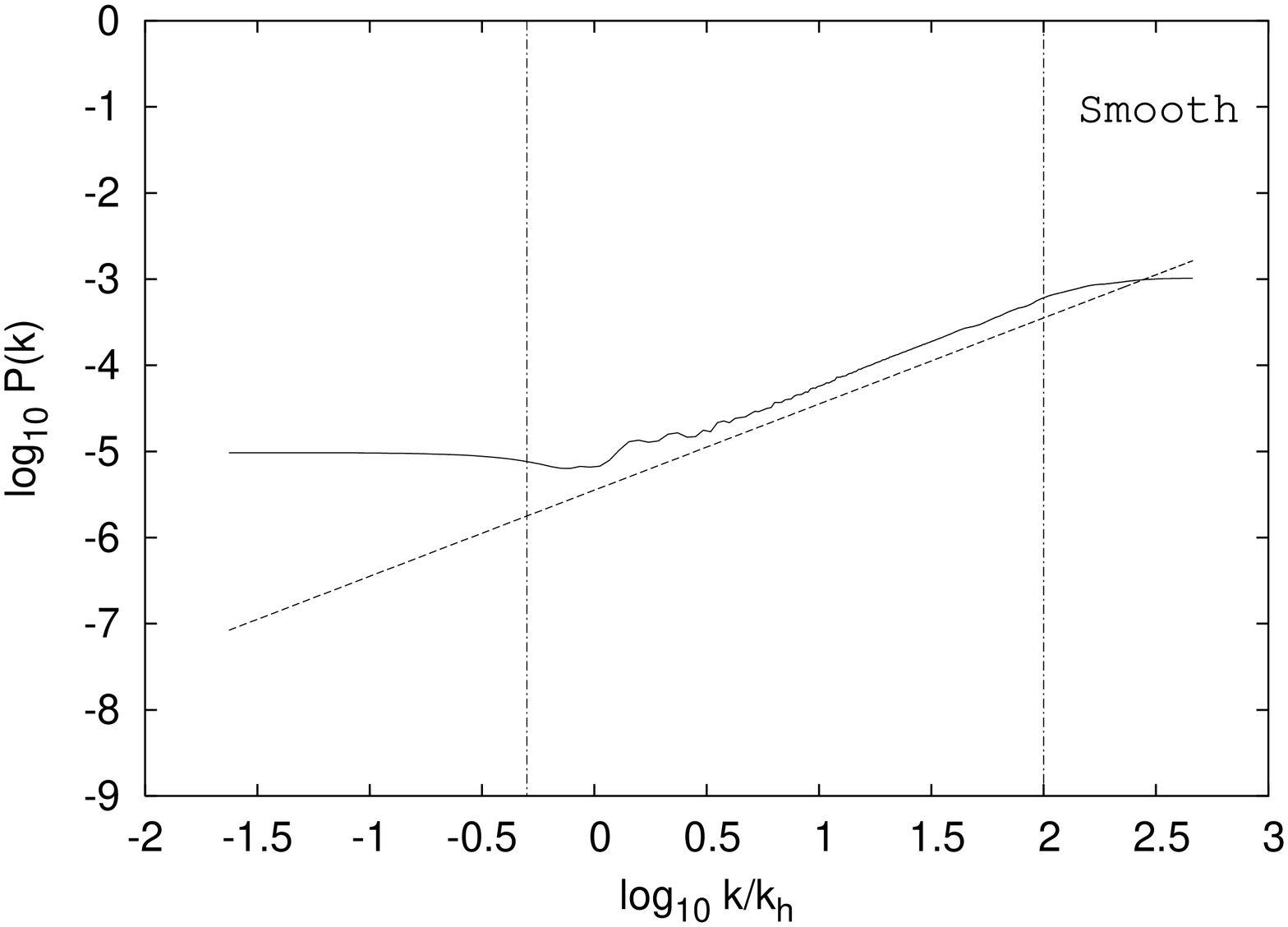}
%\epsfxsize=2.4in
%\epsffile{/dsk4/cmb/people/arman/cmb/project/phase2/continue/WMAP/plot4paper/rks1.ps} \\  

%\mbox{\bf (a')} & \mbox{\bf (b')} & \mbox{\bf (c')}
\end{array}$
$\begin{array}{@{\hspace{-0.4in}}c@{\hspace{0.3in}}c@{\hspace{0.3in}}c}
\multicolumn{1}{l}{\mbox{}} &
\multicolumn{1}{l}{\mbox{}} &
\multicolumn{1}{l}{\mbox{}} \\ [-0.5cm]

\includegraphics[scale=0.22, angle=-90]{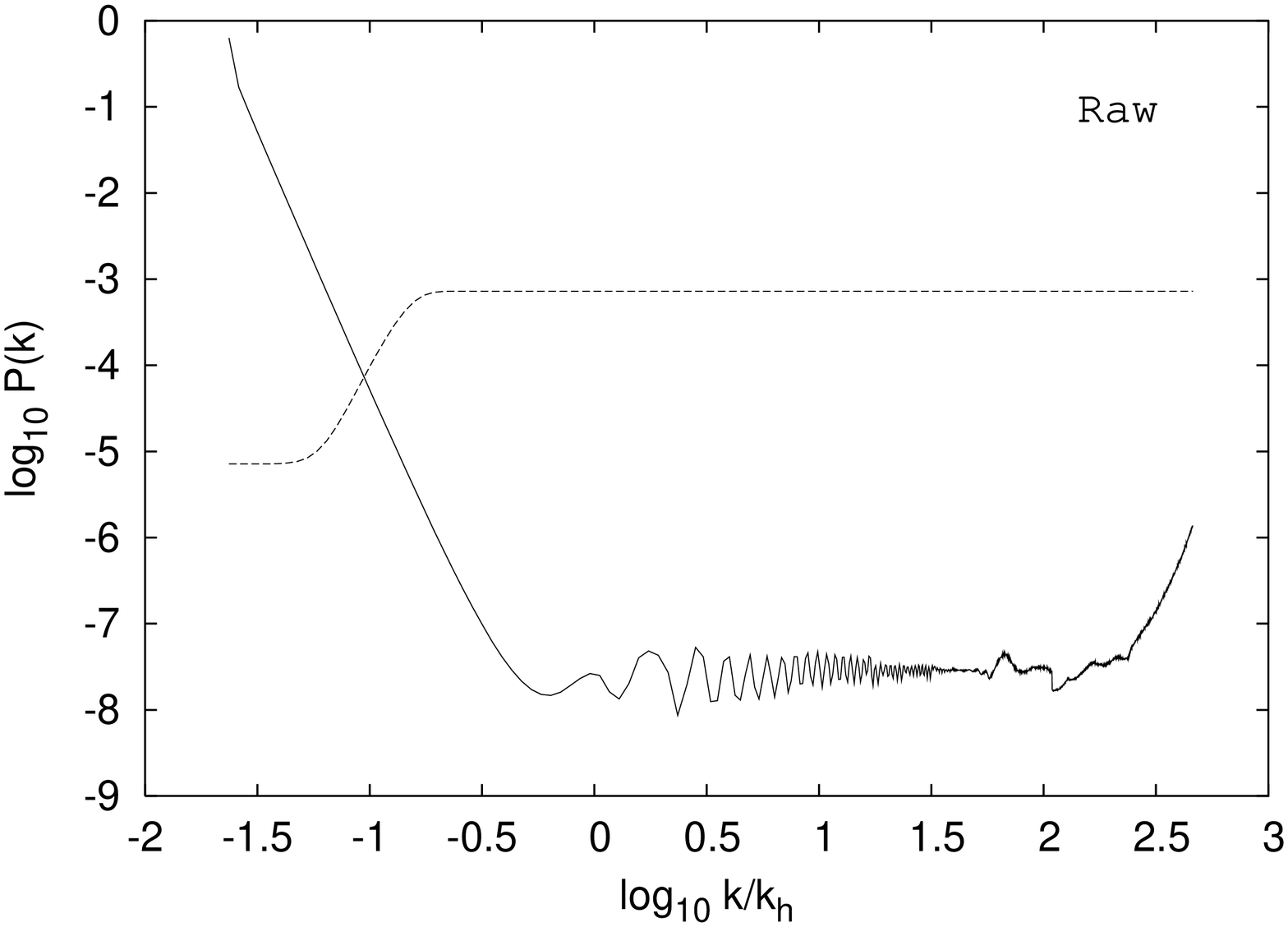}
  
\includegraphics[scale=0.22, angle=-90]{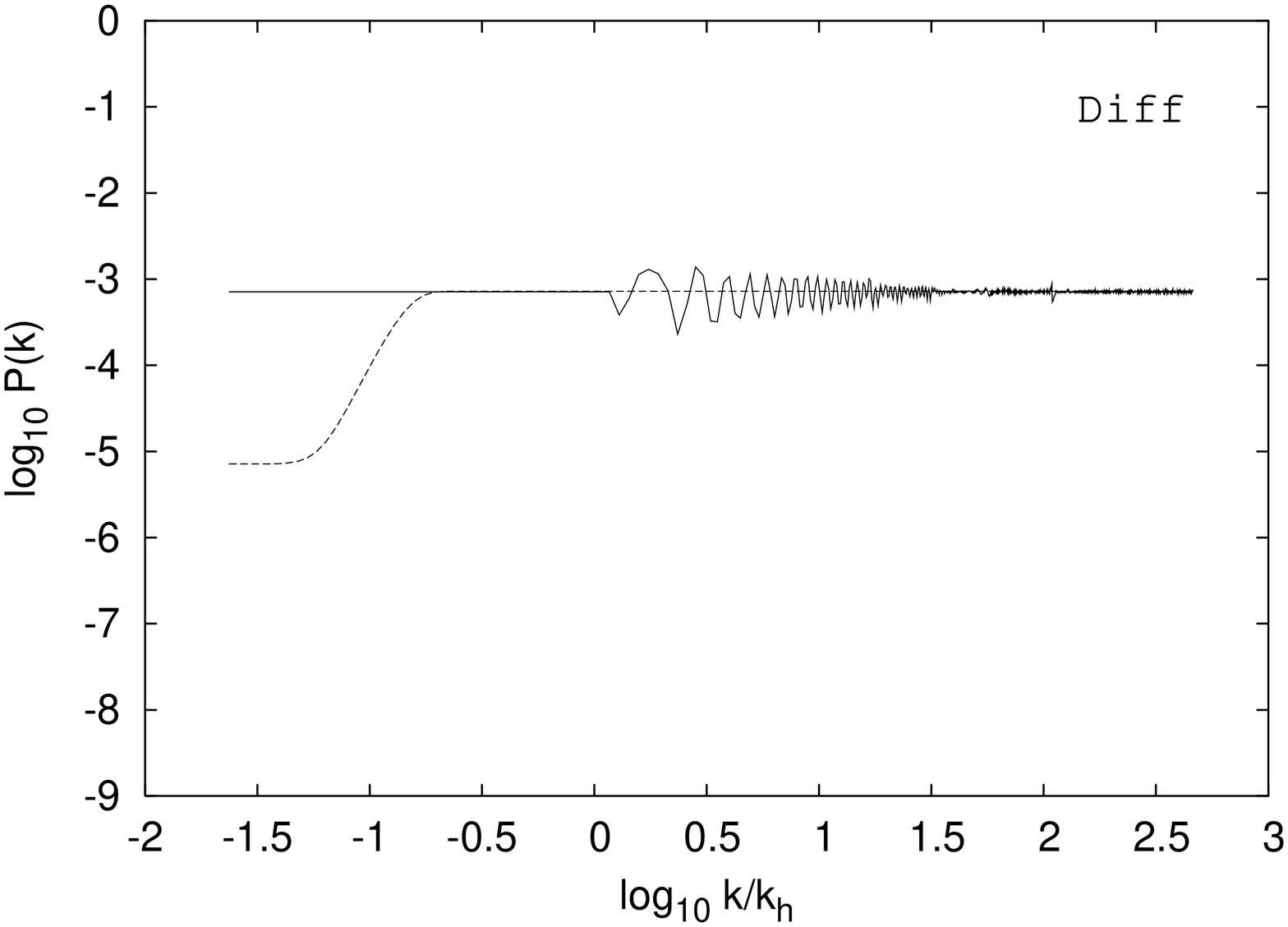}

\includegraphics[scale=0.22, angle=-90]{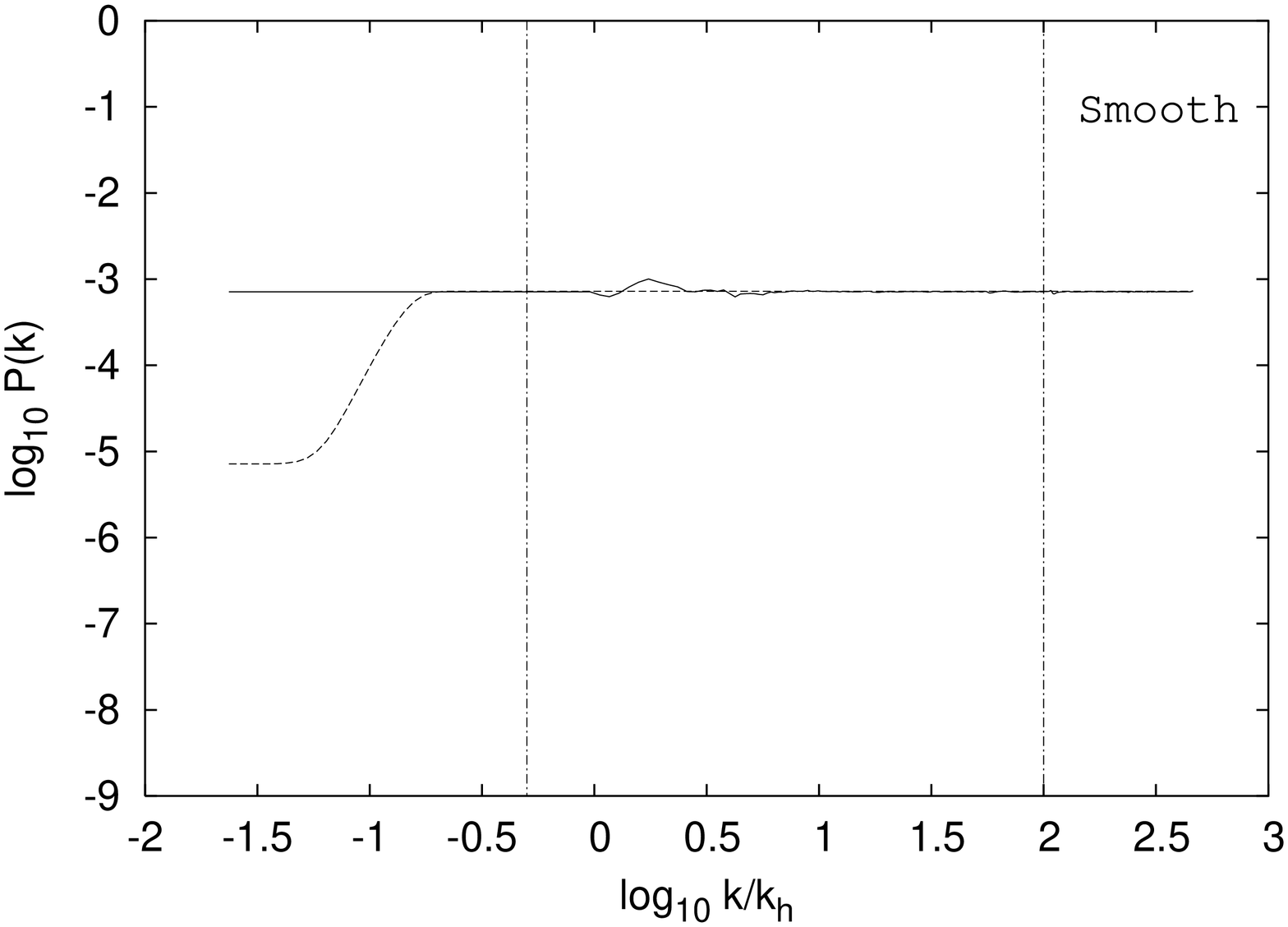}
%\epsfxsize=2.4in
%\epsffile{/dsk4/cmb/people/arman/cmb/project/phase2/continue/WMAP/plot4paper/rks1.ps} \\  

%\mbox{\bf (a')} & \mbox{\bf (b')} & \mbox{\bf (c')}
\end{array}$
\end{center}
\caption{\small Each of the three rows in the panel of figures
illustrates the recovery the primordial power for a test case using
$C_l$ arising from a known non-scale invariant primordial
spectrum. The first column compares the raw deconvolved spectrum with
the input spectrum. Note the similar artifacts in the all the raw
spectra at the low and high $k$ end discussed in the text. The feature
is outside the range of $G(l,k)$ and is completely missed in the third
case. The second, column is the differenced spectrum obtained by
dividing out by a raw reference spectrum. The differenced spectrum
resembles the input spectrum (top two cases) with small
oscillations. The third column shows that the final recovered spectrum
obtained by smoothing the differenced spectrum matched the input
spectrum very well.  }
\label{testrec}
\end{figure*}

We illustrate the steps of removing the numerical artefacts and
smoothing with test case examples of synthetic $C_l$ from known test
primordial spectrum. The first column of Fig.~\ref{testrec} shows the
`raw' spectrum obtained. A comparison with the known spectra shows
similar numerical noise and artefacts in all the cases, especially,
the rise at very low and very high wave numbers.  As shown in
appendix~\ref{refspec}, the main artifacts at the low and high $k$
ends can be understood and modeled analytically.  However we find it
easier to remove them constructing a numerically generated template
reference spectrum that also takes care of small features which appear
due to changes in the $k$ spacing.  To remove these numerical
artifacts, we generate a synthetic $C_l$ using a Harrison-Zeldovich
spectrum and then apply the deconvolution method to recover a
`reference' spectrum $P_{\rm ref}(k)$. We divide every `raw' spectrum
by $P_{\rm ref}(k)$ to obtain a `differenced' spectrum that does not
have numerical artifacts as seen in middle column (``diff.'')  of
Fig~\ref{testrec}.

The difference spectra are a noisy version of the test spectra. Hence,
the exercise with the test spectra suggests that the differenced
spectrum needs to be `suitably' smoothed to recover the true power
spectrum. A simple smoothing procedure with simple fixed width window
function leads to a satisfactory recovery of the test spectrum. (A
`Bowler hat' window constructed with hyperbolic tangent functions is
used for smoothing.)

The last column in Fig~\ref{testrec} shows the remarkably successful
recovery of the test spectra in the top two case. As expected, the
$P(k)$ is recovered best in the $k$ space (within the vertical lines
in bottom row of Fig~\ref{testrec}) where the kernel $G(l,k)$ is
significant (marked by the vertical lines in Fig.~\ref{glk} and
\ref{testrec}). In the bottom row, the feature in the test spectrum is
outside the range of the kernel $G(l,k)$, and, as expected, the
recovery process misses it completely. For the top row, if we ignore
the region where there is no power for $G(l,k)$ the recovered spectrum
is well matched with the test spectrum. The match may be further
improved by using more elaborate, adaptive smoothing procedure.

For application to real data, the phrase `suitable smoothing
procedure' may appear ambiguous. But the smoothing is in fact very
well defined for real data by demanding that the smoothed $P(k)$
produces a theoretical $C_l$ that has higher likelihood given the
data. We find that this approach works extremely well.  The
deconvolution algorithm uses the binned $C_l$ data, hence the
theoretical $C_l$ corresponding to the recovered $P(k)$ fits the
binned data very well. However, the WMAP likelihood of the theoretical
$C_l$ suffers owing to spurious oscillations in the differenced
spectra.  The WMAP likelihood improves as the differenced spectra,
$P_{\rm diff}(k)$ is smoothed.  We smooth the differenced spectrum so
that WMAP likelihood of the corresponding theoretical $C_l$ is
maximized. Although it is difficult to establish that the final result
is the unique solution with maximum likelihood, in practice, our
simple scheme does lead to a well defined result (no distinct
degenerate solutions were found). Since our smoothing procedure is
simple minded, possible avenues for improvement with more elaborate
smoothing procedure remain open. Work is in progress to employ wavelet
decomposition for the smoothing procedure.

\section{Application to the WMAP CMB anisotropy spectrum}
\label{results}

We apply the method described in the previous section to the angular
power spectrum obtained with the first year of WMAP data publicly
released in February 2003~\cite{ben_wmap03} to recover the primordial
power spectrum. In section~\ref{wmapdat}, the publicly available WMAP
data and how it is used in our work is discussed. We also describe the
choice of the `base' cosmological model.  The recovered primordial
power spectrum for this model is presented in section~\ref{primspec}.
The next section~\ref{paramvar} presents the effect of varying the
cosmological parameters (within $1~\sigma$ error bars) on the
recovered primordial power spectrum. We also present the primordial
spectrum for a set of cosmological models with large optical depths
($\tau=0.1, 0.17,0.25$) corresponding to possible the early
reionization scenarios suggested by the WMAP temperature-polarization
(TE) cross-spectrum.

\subsection{WMAP anisotropy data and the cosmological model}
\label{wmapdat}

Accurate measurements of the angular power spectrum of CMB anisotropy
was derived from the first year WMAP data
recently~\cite{hin_wmap03}. The spectrum obtained by averaging over
$28$ cross-channel power measurements is essentially independent of
the noise properties of individual radiometers. The power at each
multipole ranging from $l=20$ to $900$ was estimated together with the
covariance~\footnote{The $C_l$ are uncorrelated for an ideal full-sky
map, but in practice, the covariances between neighboring multipole
arise due to non-uniform/incomplete sky coverage, beam
non-circularity, etc..}. The instrumental errors are smaller than the
cosmic variance up to $l\sim 350$, and the signal to noise per mode is
above unity up to $l\sim 650$.

The angular power spectrum estimate and covariance matrix are publicly
available at the LAMBDA data archive~\footnote{Legacy Archive of
Microwave Background DAta -- http://lambda.gsfc.nasa.gov/}. The WMAP
team has also made available a suite of F90 codes that computes the
likelihood for a given theoretical $C_l$ spectrum given the full
angular power spectrum included the covariance measured by WMAP. We
use the TT likelihood code for computing the likelihood of $C_l$
obtained from the recovered power spectrum and refer to these numbers
as the `WMAP likelihood' in the paper.

In addition, the WMAP team has also obtained a binned angular power
spectrum where an average $C_l$ is defined over bins in multipole
space. The binned $C_l$ estimates can be treated as independent data
points since the covariance between binned estimates is negligible.
We use this binned spectrum as $C_l^D$ in the deconvolution of
eq.~(\ref{clsum}).  (We are aware of but do not consider the revised
estimates of the low multipoles made by other authors after the WMAP
results~\cite{eft03,teg03})

The variance of $C_l$ measurements is given by

\be 
\sigma_l^2 = \frac{2}{(2l+1) f_{\rm sky}}\left[ C^T_l +
\sigma_N^2\varpi_p^2 B_l^{-2}\right]^2\,,
\label{cosvar}
\ee where $\sigma_N$ is the noise per pixel, $\varpi_p$ the angular
pixel size, $f_{\rm sky}$ is the fraction of sky covered and $B_l$ is
the transform of the experimental beam~\cite{knox95}. It is important
to note that contribution to the error from cosmic variance is
proportional to the underlying theoretical/true $C_l^T$ spectrum. To
obtain the total error bars $\sigma_l$ used in the IRL deconvolution
eq.~(\ref{RLerr}), we should add the cosmic variance for the
theoretical $C_l^{(i)}$ to the the statistical error bars given with
the binned data.  However, $C_l^{(i)}$ rapidly iterates to $C_l^D$
within the error bars in the IRL method and the simpler option of
using $C_l^D$ instead $C_l^{(i)}$ for computing the cosmic variance
works equally well since $\sigma_l$ in eq.~(\ref{RLerr}) simply
regulates the convergence of IRL~\footnote{The contribution to the
error from cosmic variance given in the WMAP binned power data is
computed using the $C_l^T$ for the best fit $\Lambda$-CDM model with a
power law primordial spectrum. Our error bars are computed using
$C_l^D$.}.

We consider a flat $\Lambda$-CDM universe, with Hubble constant, $H_0
= 71~{\rm km.s^{-1}/Mpc.}$, Baryon density,
$\Omega_b\,h_0^2\,=0.0224$, a cosmological constant corresponding to
$\Omega_{\Lambda}\,=\,0.73$, with the remaining balance of matter to
critical density in cold dark matter.  This is the `concordance'
cosmological model suggested by the WMAP parameter estimation.  While
we mainly focus attention on a cosmological model without early
reionization (optical depth to reionization, $\tau=0$) we do present
in section~\ref{paramvar} the recovered primordial spectra for models
with early reionization with opacity going up to $\tau=0.25$.  The
case for a large optical depth $\tau$ comes from the
temperature-polarization cross correlation. For simplicity, we limit
ourselves to the temperature anisotropy spectrum and avoid undue
attention to the cosmology suggested by the yet incomplete
polarization data.  The polarization spectrum is expected to be
announced by WMAP soon. It is then easy to extend our method to
include both the temperature and polarization anisotropy spectra.

\begin{figure}[h]
\includegraphics[scale=0.45, angle=-90]{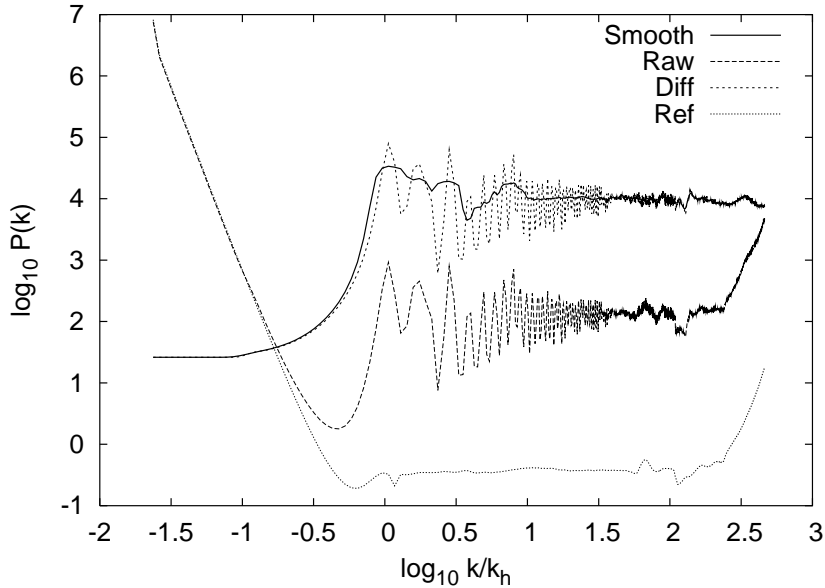}
  \caption{ The three stages leading to the final recovered spectrum
  for a base cosmological model ($\tau\,=\,0.0$, $h=0.71$,
  $\Omega_b\,h^2\,=0.0224$ and $\Omega_{\Lambda}\,=\,0.73$) is
  shown. The lower dashed line is the raw deconvolved power. The upper
  dashed line is the differenced spectrum obtained by dividing out by
  the reference spectrum shown as a dotted line. The solid line is the
  final result after smoothing that gives the best likelihood.}
\label{pk00steps}
\end{figure} 

\begin{figure*} 
\centering
\begin{center} 
\vspace{-0.05in}
\centerline{\mbox{\hspace{0.in} \hspace{2.1in}  \hspace{2.1in} }}
$\begin{array}{@{\hspace{-0.4in}}c@{\hspace{0.3in}}c@{\hspace{0.3in}}c}
\multicolumn{1}{l}{\mbox{}} &
\multicolumn{1}{l}{\mbox{}} &
\multicolumn{1}{l}{\mbox{}} \\ [-0.5cm]
 
\includegraphics[scale=0.32, angle=-90]{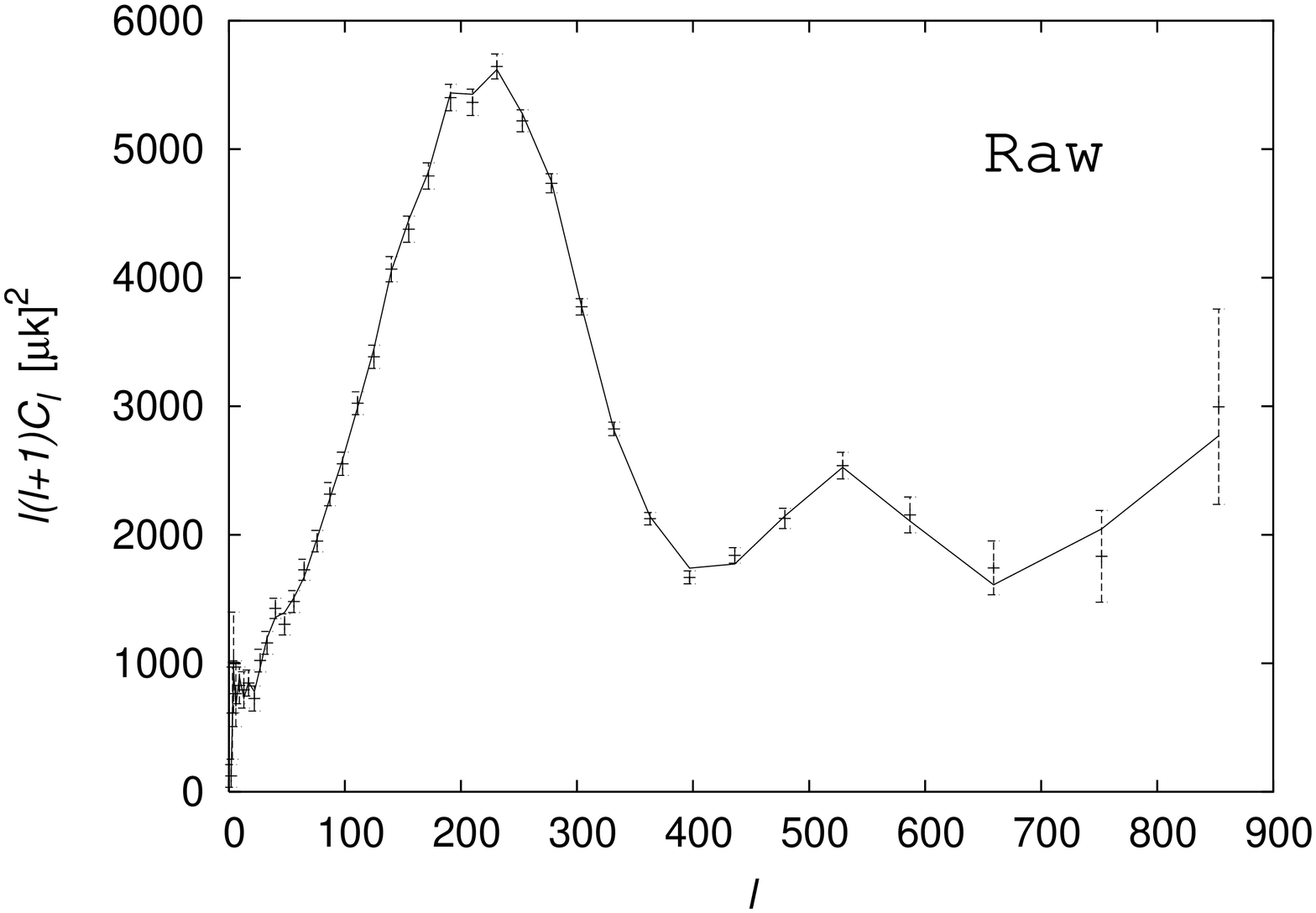}
 
\includegraphics[scale=0.32, angle=-90]{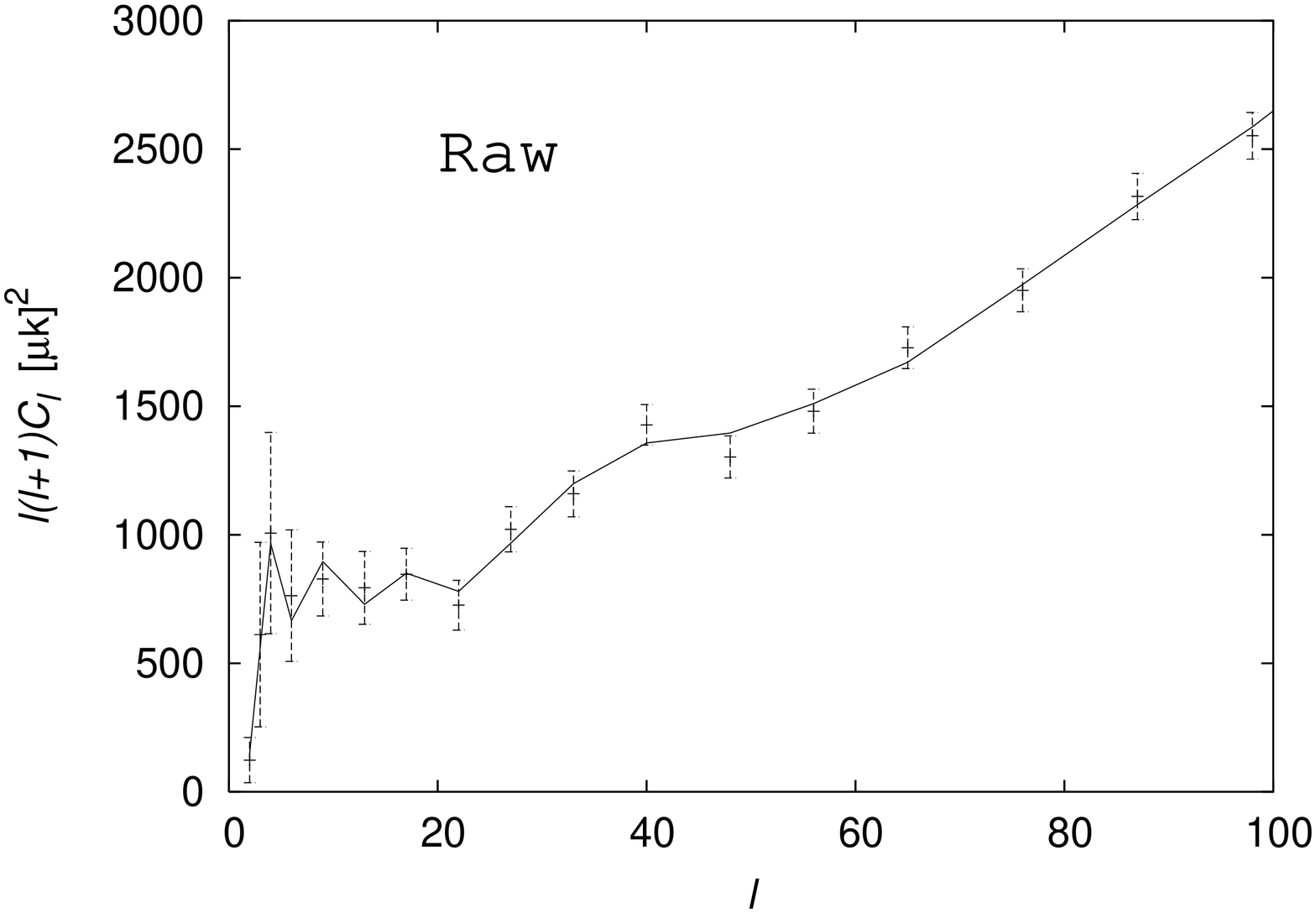}
\mbox{\bf (a)}
\end{array}$

$\begin{array}{@{\hspace{-0.4in}}c@{\hspace{0.3in}}c}
\multicolumn{1}{l}{\mbox{}} &
\multicolumn{1}{l}{\mbox{}} \\ [-0.5cm]

\includegraphics[scale=0.32, angle=-90]{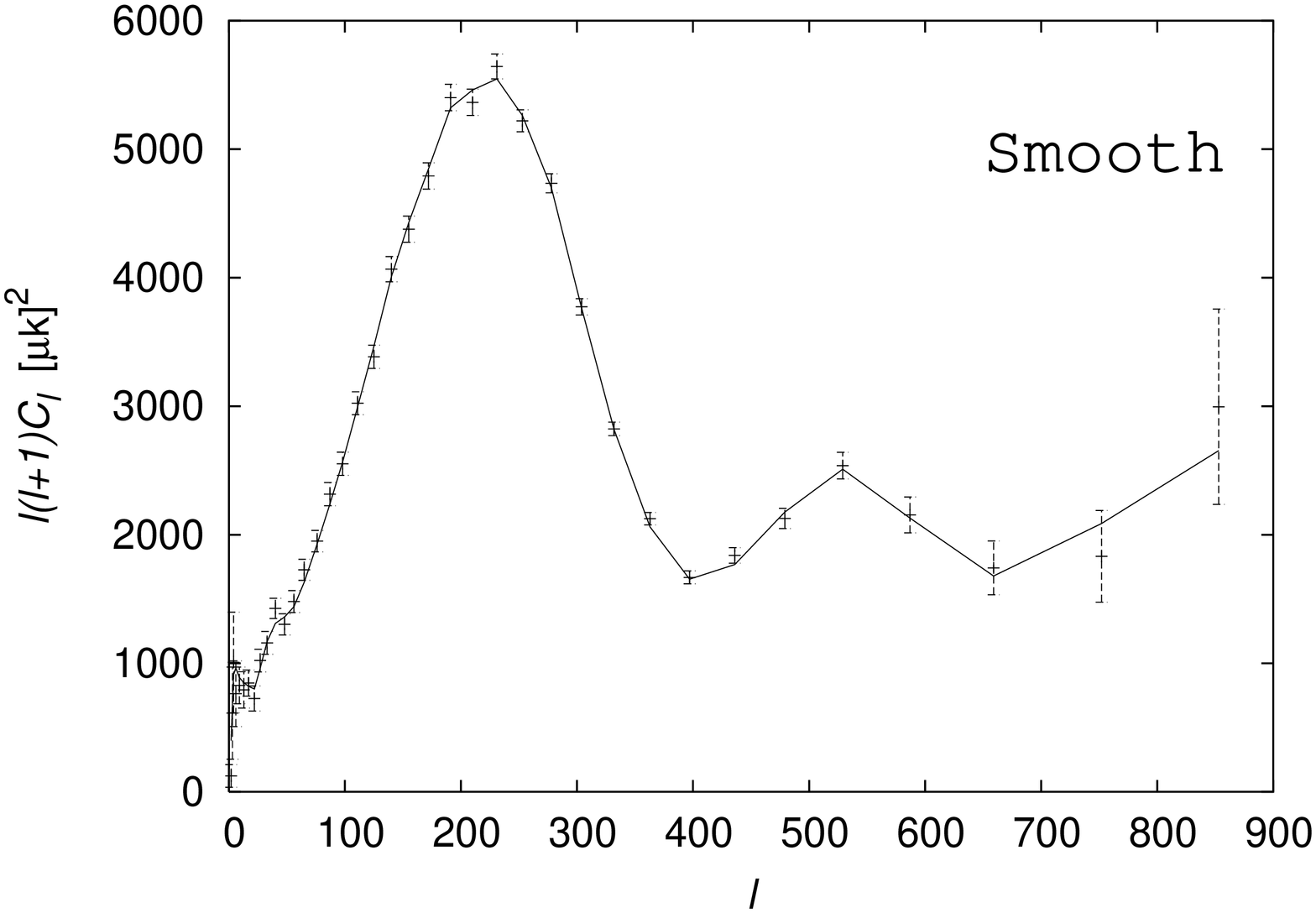}
  
\includegraphics[scale=0.32, angle=-90]{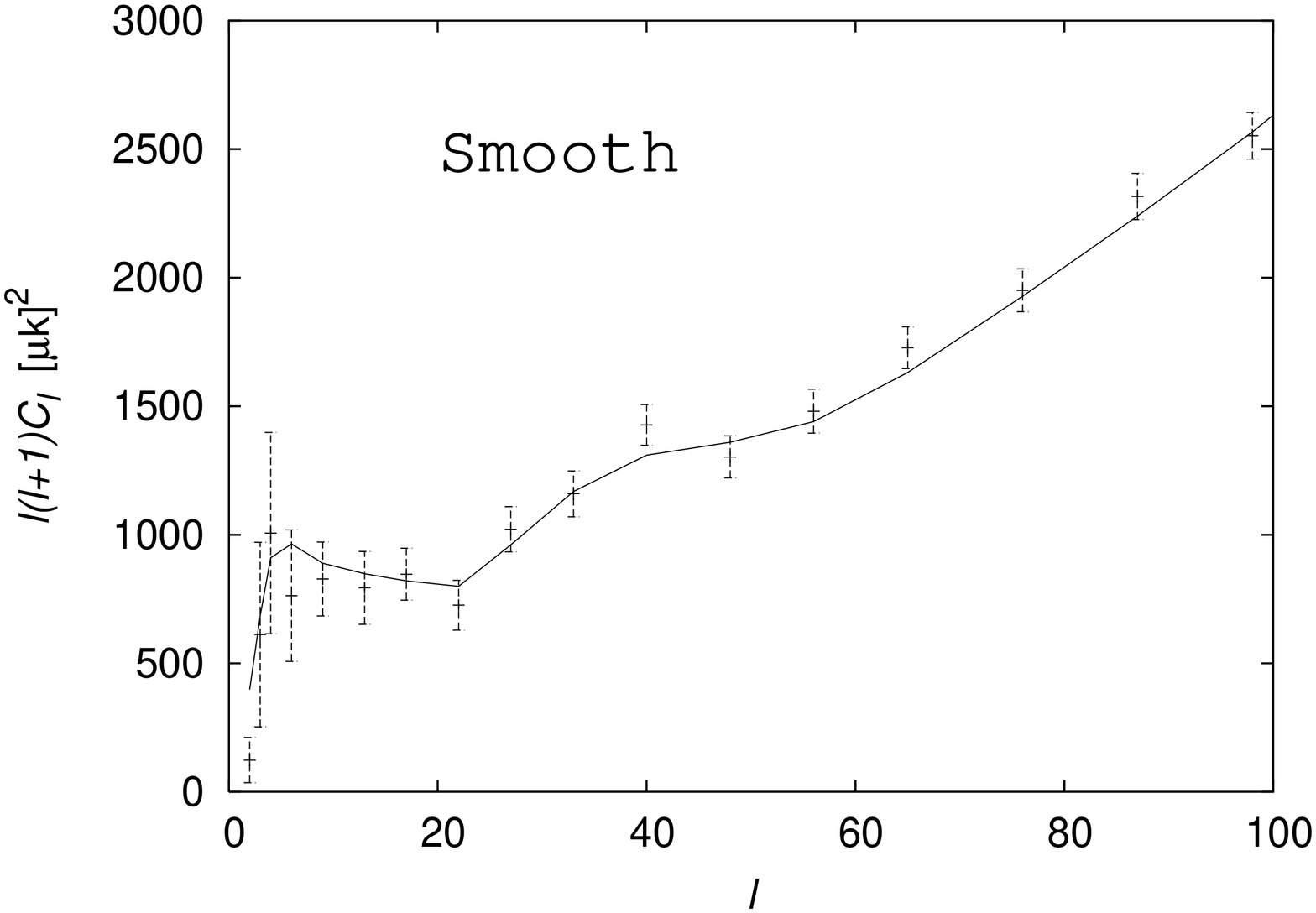}

\mbox{\bf (b)}

\end{array}$

\end{center}
\caption{\small The recovered $C_l$ corresponding to the raw $P(k)$
 are shown in the upper row and that corresponding to the final
 smoothed $P(k)$ spectrum are shown in the lower row.  The left panels
 show the full range of multipole, while the right hand panels zoom
 into the low multipoles. The $C_l$'s from the raw as well as final
 $P(k)$ fit the binned $C_l^D$ well ($\chi^2\sim 10$ and $\chi^2\sim
 20$, respectively for $38$ points). However, the jagged form of the
 $C_l$ between the $l$ bins (apparent in the low multipoles on the
 right) leads to poor WMAP likelihood for the $C_l$ from the raw
 $P(k)$. The differencing and smoothing procedure irons out the jagged
 $C_l$ dramatically improving the WMAP likelihood for the final
 smoothed $P(k)$. The WMAP likelihood is more relevant since it
 incorporates the estimation of each $C_l$ and the full error
 covariance.}
\label{cl00steps}
%%\LABEL{FIG:CL}
\end{figure*}

\begin{figure}[h]
\includegraphics[scale=0.45, angle=-90]{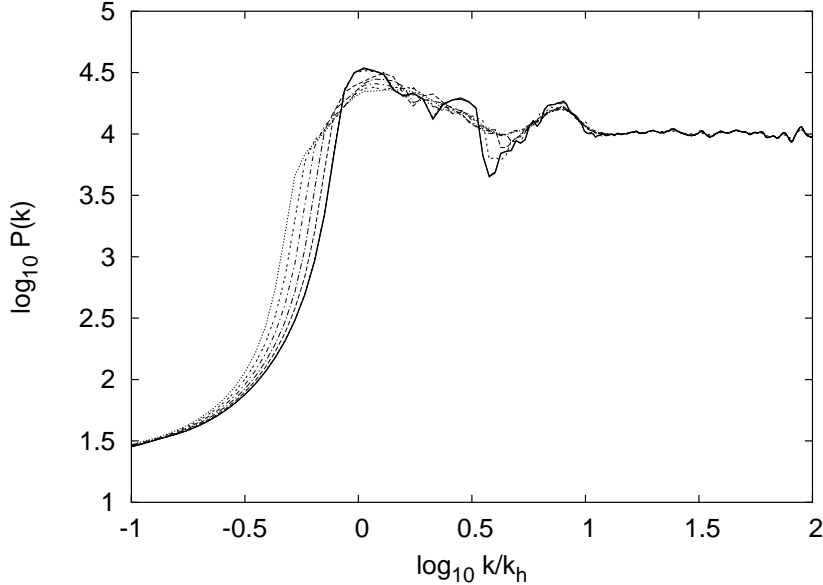}
  \caption{\small The final recovered spectrum for the base
  cosmological model ($\tau\,=\,0.0$, $h=0.71$,
  $\Omega_b\,h^2\,=0.0224$ and $\Omega_{\Lambda}\,=\,0.73$) is
  compared with set of $P(k)$ with WMAP likelihood within $\sim
  2~\sigma$. The thick line gives the best likelihood equal to
  $-478.2$ and the other lines gives the likelihood bigger than
  $-480$. We can see that the sharp infra-red cut off is common to all
  these recovered spectra. The infra-red cutoff is remarkably close to
  the horizon scale and appears to be a robust feature. Another
  significant and robust feature is the bump just above the cut-off
  (reminiscent of the oscillation from under-damped transient). The
  difference between these spectra are in the smoothing and removing
  the noises from the raw deconvolved spectrum. }
\label{pk00main}
\end{figure} 

\begin{figure}[h]
\includegraphics[scale=0.45,angle=-90]{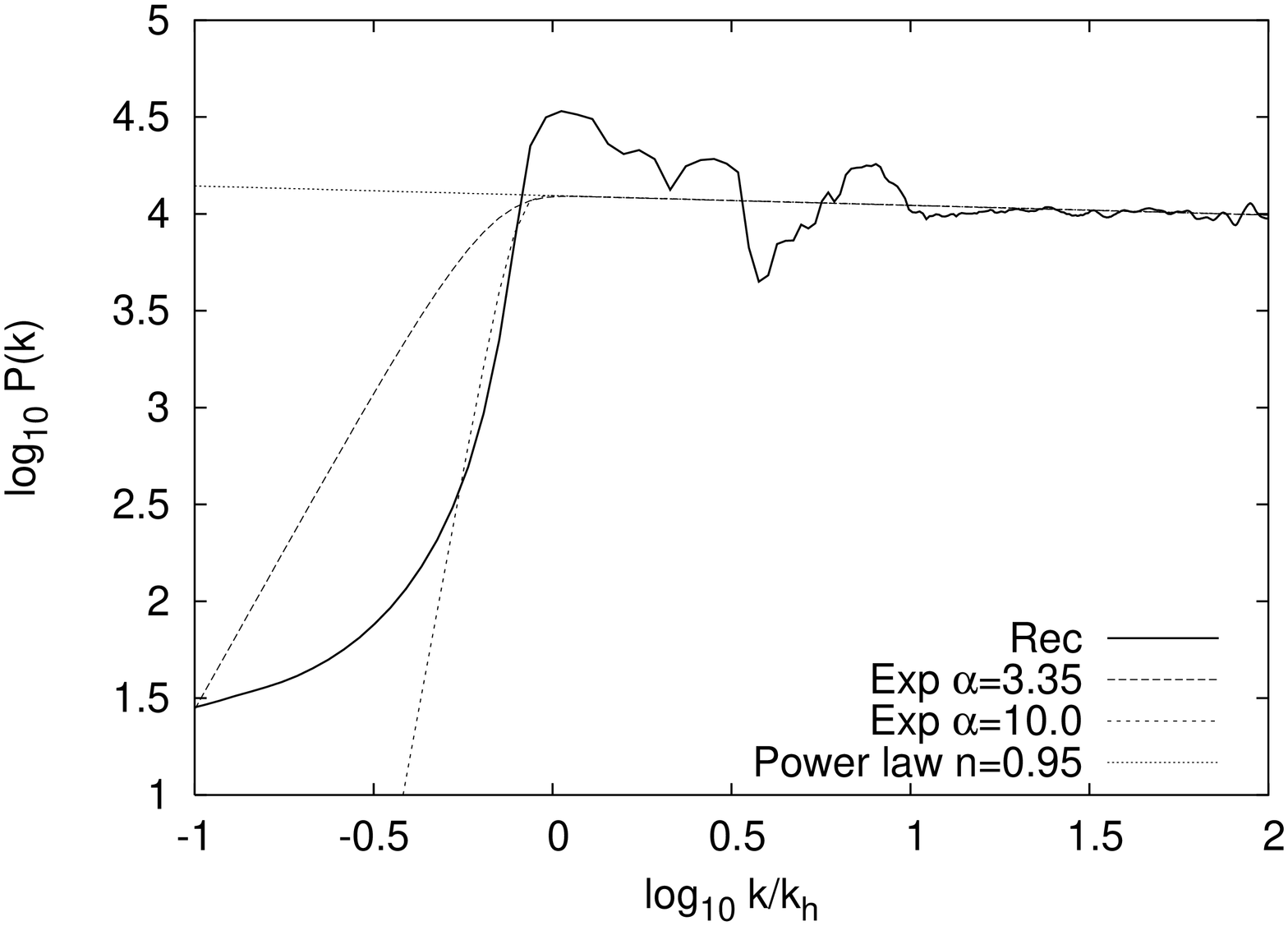}
  \caption{ Comparison between our recovered $P(k)$ and exponential
form of infra-red cut off recently studied to explain the suppressed
quadrupole of WMAP angular power spectrum~\cite{con03,clin03}. We note
that the infra-red cut-off of the recovered spectrum is very steep.
The power excess just above the cutoff in the recovered is extremely
significant in the remarkably enhanced likelihood. The power spectrum
($n=0.95$) shows that our method does recover the preferred tilt
obtained by WMAP team parameter estimation with power law
spectrum~\cite{sper_wmap03}. }
\label{pk00comp}
\end{figure}

\subsection{Primordial power spectrum from WMAP}
\label{primspec}

We apply the method described in the previous section to the WMAP
data. Fig.~\ref{pk00steps} shows the raw deconvolved spectrum, the
reference spectrum, the differenced spectrum and the final recovered
spectrum after smoothing obtained at each step of our method outlined
in section~\ref{meth}. 

We discuss the advantage of using improved Richardson-Lucy (IRL) in
appendix~\ref{irl}. The effect of IRL method is also evident in
Fig.~\ref{cl00steps}. The $C_l$ from the recovered $P(k)$ matches the
binned $C_l^D$ only within the error-bars.  The $C_l$ spectrum
corresponding to the differenced and smoothed $P(k)$ is shown in
Fig.~\ref{cl00steps}. Although, the former has better $\chi^2$ for the
binned data, the WMAP likelihood of the latter is much better.  The
poor likelihood of the differenced spectrum arises because there is no
check on fluctuations in $C_l$ at intermediate multipoles between the
bin centers arising from the spurious numerical effects of $k$-space
and $l$-space sampling. Better WMAP likelihood is more relevant than
good $\chi^2$ for the binned data since the former incorporates the
estimation of each $C_l$ and the full error covariance.  Thus the
smoothing step our method is carried out with a well defined goal of
maximizing the WMAP likelihood.

The primordial spectra recovered from WMAP data is again shown in
Fig.~\ref{pk00main}. The dark solid line is the primordial spectrum
that has the best WMAP log-likelihood of $\ln\,{\cal L} = -478.20$ for
the base cosmological model described in the previous section. The
other lines have likelihood $\ln\,{\cal L}> -480$, i.e., roughly
within $2~\sigma$ of the best one. For comparison, the same
cosmological model with a scale invariant (Harrison-Zeldovich)
primordial spectrum has $\ln\,{\cal L}= -503.6$, and, with a tilted
scale free primordial spectrum $n=0.95$, the likelihood improves to
$\ln\,{\cal L}= -489.3$. (A comparison of likelihood numbers for
various primordial spectra is given in Table~\ref{thlike}.) The
improvement in likelihood for the recovered spectra is striking.

The most prominent feature of the recovered spectrum is the infra-red
cutoff in the power spectrum remarkably close to the horizon scale
($k_c\sim k_h\equiv 2\pi/\eta_0$) chiefly in response to the low
quadrupole measured by WMAP. Another notable point is the slight tilt
($n\sim 0.95$) of the plateau in the recovered spectrum at large $k$
which is consistent with the best fit $n$ obtained by WMAP for power
law primordial spectra.  After the WMAP results, model power spectra
with infra-red cutoffs of the form

\be
P(k)\,=\,A_s\,k^{1-n_s}\,\left[1-e^{-(k/k_*)^{\alpha}}\right] 
\label{expcut}
\ee invoked to explain the suppressed low multipoles of WMAP have had
limited success~\cite{con03,clin03}.  While the effective $\chi_{\rm
eff}^2\equiv -2 \ln{\cal L}$ improved by at least $22$ over a power
law model, the model infra-red cutoff eq.~(\ref{expcut}) could improve
the $\chi_{\rm eff}^2$ merely by $\sim 3$~\cite{clin03}.
Figure~\ref{pk00comp} compares the recovered spectrum with these model
infra-red spectra. The power law spectrum shown in the figure also
highlights the small tilt ($n\sim 0.95$) recovered by our method.

\begin{figure*} 
\centering
\begin{center} 
\vspace{-0.05in}
\centerline{\mbox{\hspace{0.in} \hspace{2.1in}  \hspace{2.1in} }}
$\begin{array}{@{\hspace{-0.4in}}c@{\hspace{0.3in}}c@{\hspace{0.3in}}c}
\multicolumn{1}{l}{\mbox{}} &
\multicolumn{1}{l}{\mbox{}} &
\multicolumn{1}{l}{\mbox{}} \\ [-0.5cm]
 
\includegraphics[scale=0.32, angle=-90]{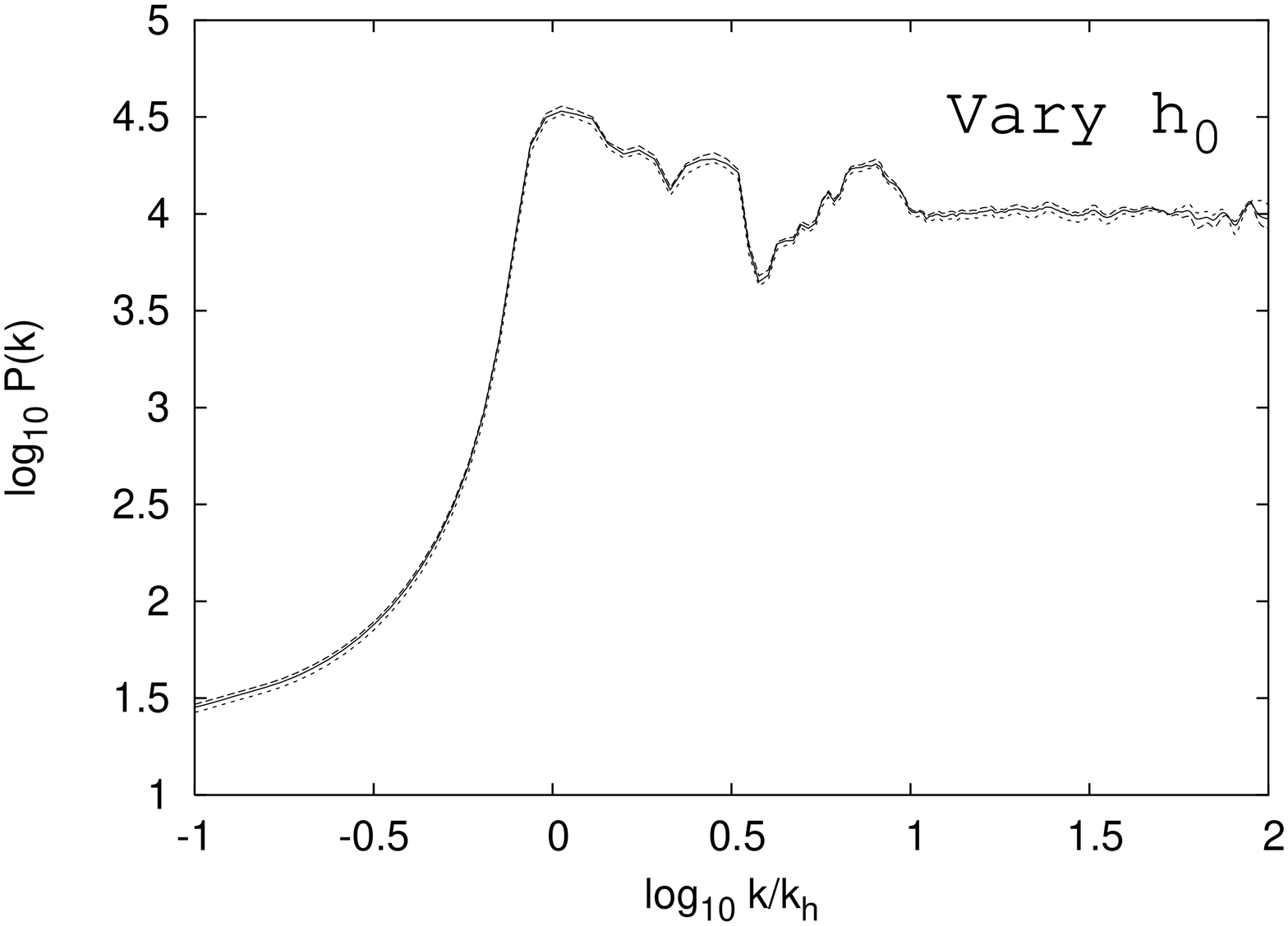}
\mbox{\bf (a)}
\includegraphics[scale=0.32, angle=-90]{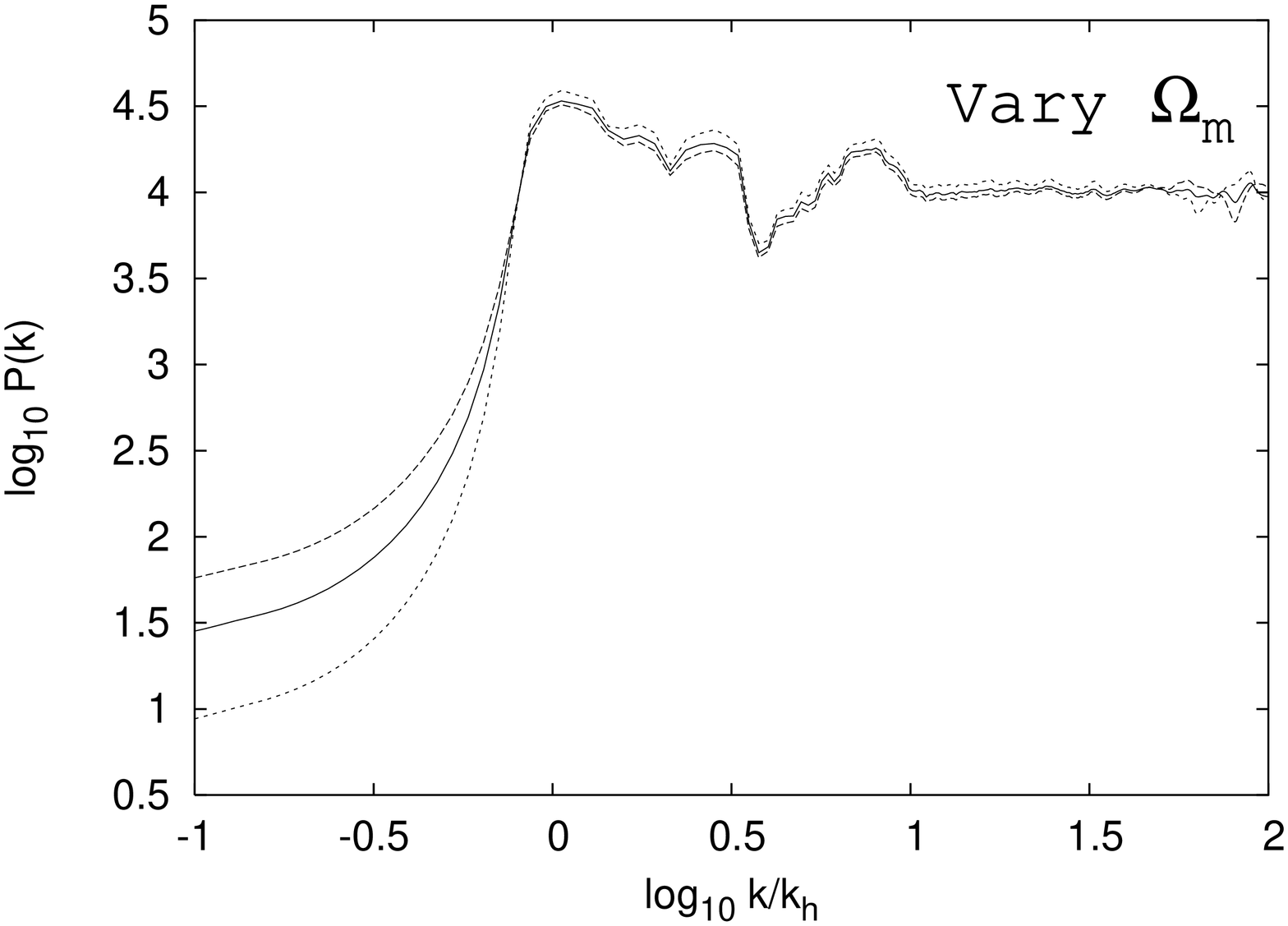}
\mbox{\bf (b)}
\end{array}$

$\begin{array}{@{\hspace{-0.4in}}c@{\hspace{0.3in}}c}
\multicolumn{1}{l}{\mbox{}} &
\multicolumn{1}{l}{\mbox{}} \\ [-0.5cm]

\includegraphics[scale=0.32, angle=-90]{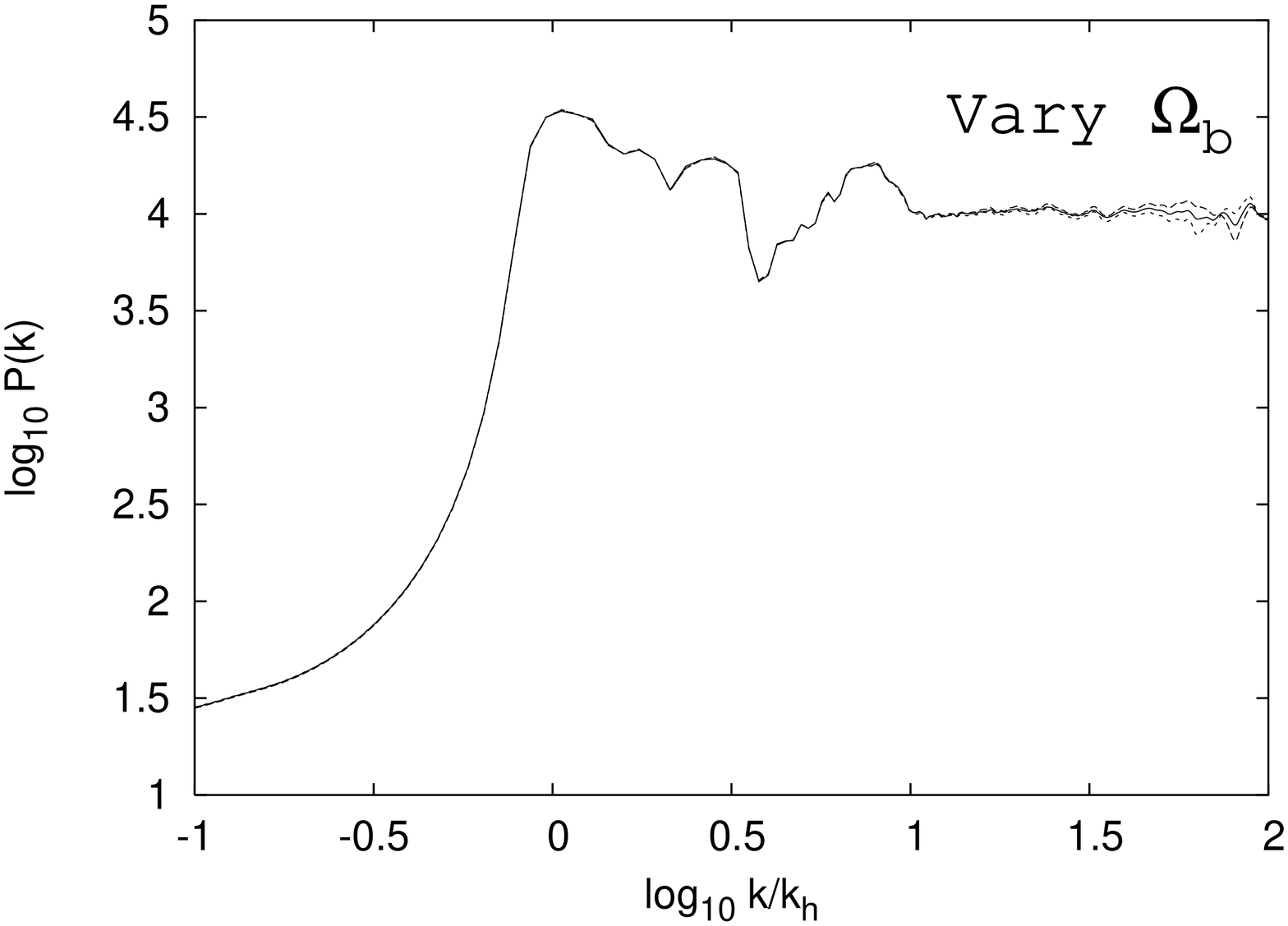}
\mbox{\bf (c)}  

\includegraphics[scale=0.32, angle=-90]{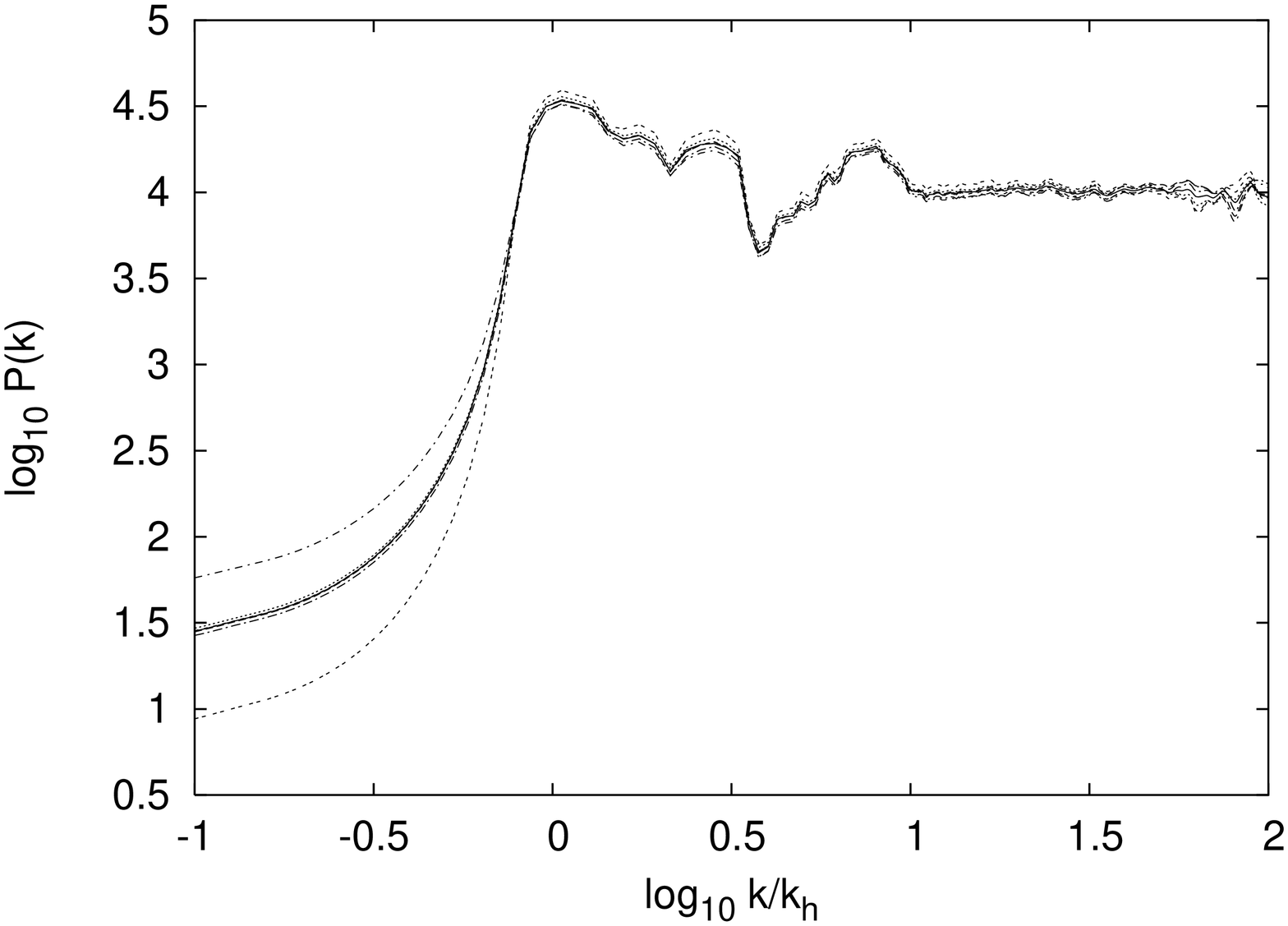}

\mbox{\bf (d)}
\end{array}$

\end{center}
\caption{\small The panel of figures shows the robustness of the
recovered $P(k)$ for variations in the cosmological parameters. Each
parameter is varied within the $1~\sigma$ range indicated by the WMAP
parameter estimates~\cite{sper_wmap03}. In Fig. (a), the Hubble
constant $h_0=0.68$, $h_0=0.71$ and $h_0=0.75$ in the three curves. In
Fig $(b)$ the values of vacuum density $\Omega_{\Lambda}\,=\,0.69$,
$\Omega_{\Lambda}\,=\,0.73$ and $\Omega_{\Lambda}\,=\,0.77$ in the
three curves. In fig.~(c), the Baryonic density $\Omega_B\,=\,0.040$,
$\Omega_B\,=\,0.044$ and $\Omega_B\,=\,0.048$ in the three curves.
The fig. (d) combines all the distinct curves in other figures to give
a consolidated perspective on the dependence of the recovered spectrum
on cosmological parameters. Note that the $x$-axis is the wavenumber
is scaled in units of the $k_h=2\pi/\eta_0$ which reduces the scatter
considerably in the curves for variations in $H_0$ and
$\Omega_\Lambda$.}
\label{vps}
\end{figure*}

\begin{figure}[h]
\includegraphics[scale=0.45, angle=-90]{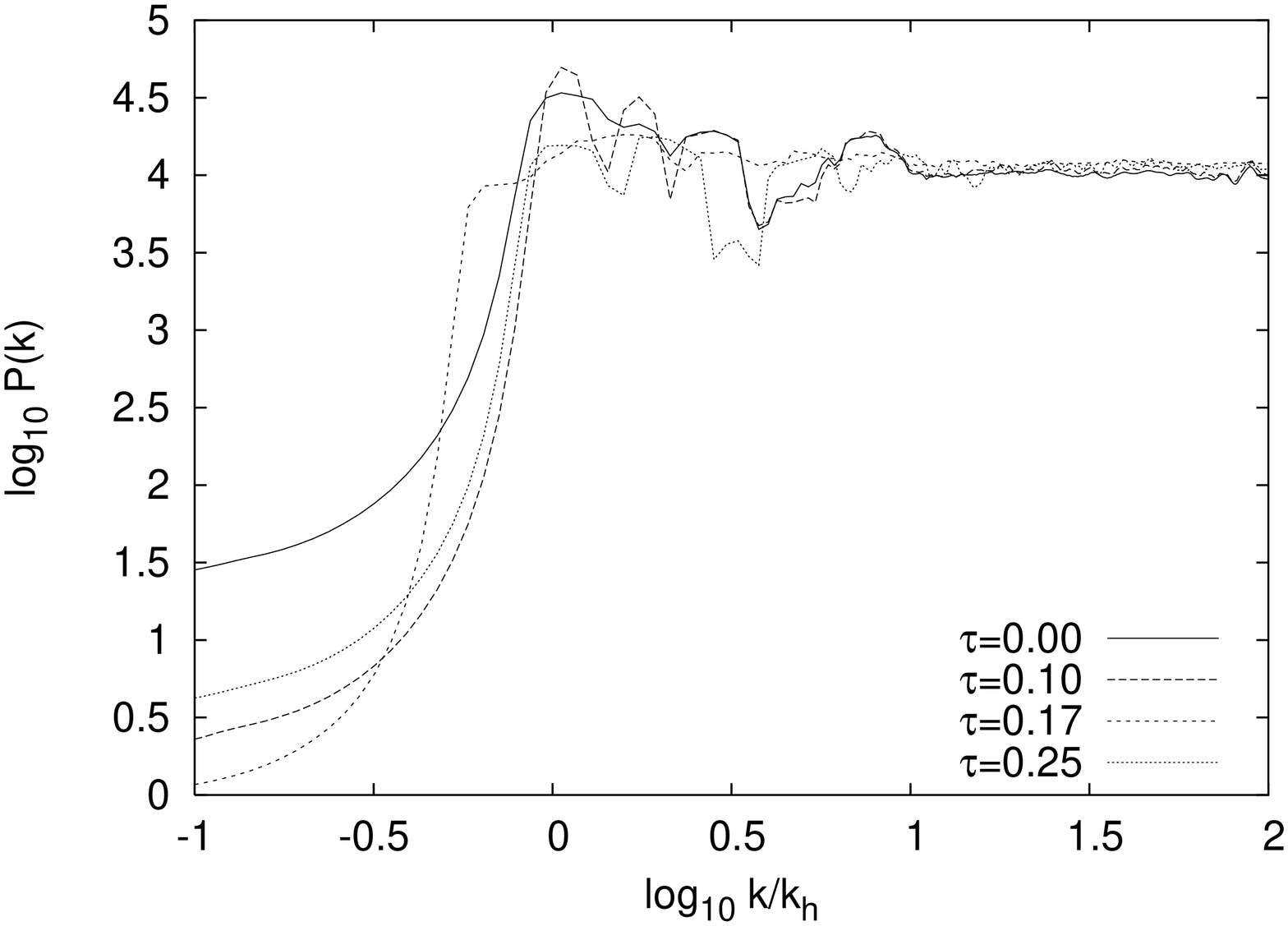}
  \caption{ The recovered $P(k)$ for different values of optical depth
  ($\tau =0.00$, $0.10$, $0.17$ and $0.25$). The width of smoothing
  used is different for different cases. This is the main cause of the
  small shift in the location of the infra-red cutoff.  }
\label{tau}
\end{figure}

\subsection{Dependence on the cosmological parameters}
\label{paramvar}

The primordial spectrum is recovered for a set of `best fit'
cosmological parameters defining the `base' model. We have varied each
cosmological parameter by the quoted $1~\sigma$ error-bars of the WMAP
estimates~\cite{sper_wmap03}.  Figure~\ref{vps} shows the dependence
of the recovered power spectrum on small changes to cosmological
parameters varied one at a time keeping the others fixed at their
central value.  We find that recovered spectrum is insensitive to
variations Hubble constant, $H_0$, modulo an overall shift due to the
change in the horizon size, $\eta_0$. The WMAP likelihood of the
corresponding $C_l$ are $\ln{\cal L} = -477.79$ and $ -478.83$ at
$h_0=0.68$ and $0.75$, respectively.  The variation of Baryon density
around its middle value $\Omega_B\,=\,0.044$ yields almost identical
primordial power spectra with a minor change in likelihood --
$\ln{\cal L} =-477.97$ and $-478.53$ for $\Omega_B\,=\,0.040$ and
$\Omega_B\,=\,0.048$, respectively.  The variation of
$\Omega_{\Lambda}$ (or equivalently, $\Omega_m$, for flat models)
modulo a shift due to change in $\eta_0$, affects only the amplitude
of the infra-red cut-off.  The WMAP likelihood of the corresponding
$C_l$ are $\ln{\cal L} = -478.93$ and $ -478.25$ for
$\Omega_{\Lambda}\,=\,0.69$ and $\Omega_{\Lambda}\,=\,0.77$,
respectively.

We note that the likelihood at the central values are not necessarily
the best. This suggests that, as discussed in the introduction, one
should explore the entire space of cosmological parameters and compute
the likelihood after optimizing with `best' recovered primordial
spectrum. Such an analysis is certainly possible, but has large
demands on computational resources. Hence, we defer it to a future
publication.

Finally, we also consider the effect of a cosmological model with
significant optical depth to reionization, $\tau \,=\,0.17$, such as
suggested by the angular power spectrum of the
temperature-polarization ($TE$) cross correlation from
WMAP~\cite{kog_wmap03}. It has been pointed out that the estimated
$\tau$ increases when the primordial spectrum has an infra-red
cut-off. We compute the primordial power spectrum for base
cosmological models with $\tau \,=\,0.10,\,0.17,\, 0.25$.  The
standard RL deconvolution performs poorly for these models, and, the
improved RL method was crucial for these models.  Fig.~\ref{tau}
compares the recovered primordial spectra for early reionization
models which all show an infra-red cutoff at the horizon
scale~\footnote{ For simplicity and consistency, we use the same
smoothing for these cases. Better result can be obtained for $\tau =
0.17$ with different smoothing.}.

\section{Theoretical implications of the recovered spectra }
\label{disc}   

The direct recovery of the primordial power spectrum has revealed an
infra-red cutoff of a very specific form. Model spectra with monotonic
infra-red cutoff such as that in eq.~(\ref{expcut}) do not improve the
WMAP likelihood significantly. While, to match the low value of the
quadrupole, a very sharp cut-off (such as $\alpha\sim 10$, see
fig~\ref{pk00comp}) is required, such a steep monotonic cut-off tends
to pull down the power in the next few higher multipoles above the
quadrupole and octopole as well. Our recovered spectrum has a
compensating excess which allows a steep cut-off to match the low
quadrupole and octopole without suppressing the higher multipoles.
Naively, one would think that a designer infra-red cutoff would `cost'
in the language of Bayesian evidence due to the introduction of extra
parameters. That is is not necessarily so. An infra-red cut-off of the
form we recover does not necessarily have more parameters than an
infra-red cut-off of the form in eq.~(\ref{expcut}). Moreover, it is
striking that the location of the cut-off is close to a well known
scale -- the horizon scale.

\begin{figure}[h]
\includegraphics[scale=0.45, angle=-90]{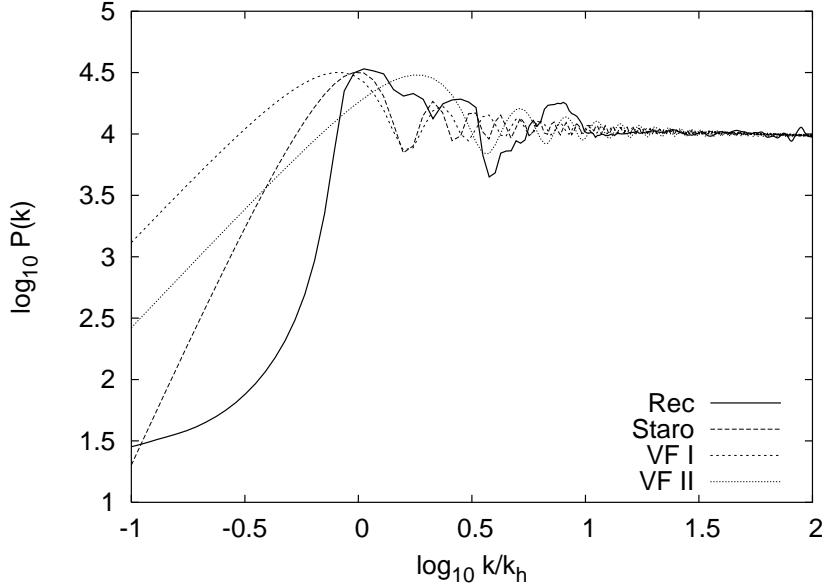}
\caption{Comparison of our recovered $P(k)$ (solid) with the
predictions two simple theoretical scenarios that remarkably match the
gross features of the infra-red cutoff in the recovered spectrum. The
`staro' curve is the primordial spectrum when the inflaton potential
has a kink-- a sharp, but rounded, change in slope~\cite{star92}. Fine
tuning is involved in locating the kink appropriately. The `VF' curves
are the modification to the power spectrum from a pre-inflationary
(here, radiation dominated) epoch~\cite{vilfor82}. This requires that
the horizon scale, $k_h$, exits the Hubble radius very soon after the
onset of inflation. Although, it appears fine tuned there is
corroborating support for this within single scalar field driven
inflation~\cite{las_sour03}.  The theoretical $P(k)$ leads to $C_l$
that enhanced WMAP likelihood given in Table~\ref{thlike}. The values
of the parameters for the theoretical curves are given in the same
table. }
\label{thpk}
\end{figure}

In this section we show that infra-red cut-off of the form we recover
arises from very simple scenarios in inflation. We explicitly mention
two of them. Starobinsky~\cite{star92} has shown that a kink (sharp
change in the slope) in the inflaton potential can modulate the
underlying primordial power spectrum $P_o(k)$ with a step like feature
at a wavenumber $k_c$

\bea P(k) &=& P_o(k) D(k,k_c,r) \nonumber \\
\,&=&\,A_s\,k^{1-n_s}\,\,[1-3(r-1)\frac{1}{y}((1-\frac{1}{y^2})\sin2y+\frac{2}{y}\cos2y) \nonumber \\ &+&\frac{9}{2}(r
-1)^2\frac{1}{y^2} ((1+\frac{1}{y^2})\cos2y-\frac{2}{y}\sin2y) ]
\label{staro}
\eea where we assume a power law $P_o(k)= A_s\,k^{1-n_s}$,
$y={k}/{k_c}$ and $r$ is the ratio of the slope $dV/d\phi$ before and
after kink in the inflaton potential. The power spectrum $P(k)$ in
eq.~(\ref{staro}) has a step up (going to larger $k$ ) for $r<1$ and a
step down feature for $r>1$.  An infra-red cut off is created with
$r<1$. Fig.~\ref{thpk} shows a spectrum with a Starobinsky step
(eq.~\ref{staro}) that can not only mimic the sharp infra-red cutoff
but also produces the required bump after it. Table~\ref{thlike} shows
that a introducing an appropriate Starobinsky step gives a very good
WMAP likelihood compared to eq.~(\ref{expcut}). Besides the location
of the break, $k_c$, the Starobinsky step spectrum has only another
parameter $r$ that fixes the slope and the depth of the cut-off as
well as the size of the bump. We have not systematically searched
through the $\{r, k_c\}$ parameter space to arrive at a `best-fit'
model. Hence, it may be possible to get even better match to the WMAP
data with Starobinsky breaks. Similar scenarios have been studied
earlier~\cite{leac01} and has also been pointed to in the post-WMAP
literature~\cite{kal_kap03}. Multiple scalar field inflation provide
ample scope for generating features in the primordial
spectrum~\cite{oldbsi} and has been been invoked to model a sharp
cut-off at horizon scale (see eg.,~\cite{yok99,fen_zhan03}).  More
exotic origin of an infra-red cut off in the scalar spectrum have also
been
investigated~\cite{tsuj_maar03,tsuj_sin03,fuk_kon03,piao_fen03,piao_tsuj03}.

Another compelling theoretical scenario for generating a feature of
the form we have recovered is well-known. It is well known that
radiation, or matter dominated era prior to inflation does affect the
primordial power spectrum on scales that `exit the horizon' soon after
the onset of inflation.  For a pre-inflationary radiation dominated
epoch the power spectrum was given by Vilenkin and Ford
(VF)~\cite{vilfor82}

\be 
P(k) = A_s\,k^{1-n_s}\,\frac{1}{4 y^4}\left| e^{-2 i y} (1+2 i y)
-1 - 2 y^2\right|^2
\label{vf}
\ee where $y={k}/{k_c}$. Fig.~\ref{thpk} shows that the VF spectrum
(eq.~\ref{vf}) can also provide an infra-red cutoff with required bump
after it. Table~\ref{thlike} shows that a VF spectrum can give better
WMAP likelihood compared to model spectra of the form
eq.~(\ref{expcut}).  The infra-red cut-off ($\propto k^2$) here is not
very sharp. However, if the epoch prior to inflation is dominated by
matter with some other equation of state, the slope would be
different.  A more complete analysis may give rise to spectra closer
to the kind we have recovered.  The scale $k_c$ is set by the Hubble
parameter at the onset of inflation. For this scenario to applicable
to our results, the $k_h$-mode corresponding to the horizon scale must
have crossed the Hubble radius very close to the onset of inflation.
In a single scalar field inflation this would happen
naturally~\cite{las_sour03}.

Another possibility is that the infra-red cutoff arises due to
non-trivial topology of the universe. A dodecahedral universe model
that matches the low multipoles of WMAP angular spectrum has been
proposed~\cite{lum03}. In future work we would like to check whether
the recovery of a discrete initial spectrum with our method appears
similar to the spectrum we recovered here. However, non-trivial cosmic
topology is expected to also violate the statistical isotropy of the
CMB anisotropy and give rise of correlation features which are
potentially detectable~\cite{haj_sour03}.

\section{Discussion and conclusions}

The CMB anisotropy is usually expected to be statistically isotropic
and Gaussian~\cite{mun_sour95}. In that case, the angular power
spectrum of CMB anisotropy encodes all the information that may be
obtained from the primary CMB anisotropy, in particular, the
estimation of cosmological parameters.  It is very important to note
that cosmological parameters estimated from CMB anisotropy (and other
similar observations of the perturbed universe) usually assume a
simple parametric form of spectrum of primordial perturbations. It is
clear, however, that estimation of cosmological parameters depends on
the extent and nature of parameterization of the primordial (initial)
perturbations included into the parameter space
considered~\cite{uscosmo97}.

We proceed on a complimentary path of determining the primordial power
spectrum directly from the CMB anisotropy for a set of cosmological
parameters. Assuming the best fit cosmological parameters from WMAP,
our method applied to the angular power spectrum measured by WMAP
yields interesting deviations from scale invariance in the recovered
primordial power spectrum.  The recovered spectrum shows an infrared
cut-off that is robust to small changes in the assumed cosmological
parameters. The recovered spectrum points to the form of infra-red cut
off that matches the low multipoles of WMAP. We also show that the
such forms of infra-red cut-off can arise from simple well-known
effects with inflation. It is important to recall that the angular
power spectrum from the `full' sky CMB anisotropy measurement by
COBE-DMR~\cite{smoot92} also indicated an infra-red
cutoff~\cite{jin_fan94}. Although we mostly emphasize the infrared
cutoff, the final recovered spectrum shows a damped oscillatory
feature after the infra-red break (`ringing'). It has been pointed out
recently that such oscillations improve $\chi^2_{\rm eff}$ and could
be possibly a signature of trans-Planckian
effects~\cite{kog03,mar_rin03}.  We have not assessed the significance
and robustness of the features at large $k/k_h$ that may be consistent
with features deduced in the analysis of recent redshift
surveys~\cite{toc_dous_sil04}.

\begin{table} 
\caption{The effective chi-square, $\chi_{\rm eff}^2 \equiv -2
\ln{\cal L}$, of the $C_l$ corresponding our recovered spectrum is
compared with a number of model primordial spectrum (with or without
the infra-red cutoff.  Limited attempt has been made to search for the
best parameter values and the $\chi_{\rm eff}^2$ for the model spectra
should be treated as indicative and are strictly upper bounds.  }
\begin{center}
\begin{tabular}{lcc}
\hline

Power spectrum & $\chi_{\rm eff}^2\equiv -2 \ln{\cal L}$ & $k_c/k_h$\footnotemark\\
&&($k_h=4.5\times 10^{-4} {\rm Mpc.}^{-1})$ \\
\hline
Direct Recovered & 956.76  & 0.71 \\
\hline
Flat Harrison Zeldovich & 1007.28 & -- \\
Power Law & 978.60 & --\\
($n_s=0.95$) && \\
Exponential cutoff  & 978.08 & 0.64  \\
($n_s=0.95$, $\alpha=3.35$) &&   \\
Exponential cutoff  & 977.84 & 0.64 \\
($n_s=0.95$, $\alpha=10$) & \\
Starobinsky break & 973.86 & 0.32 \\
 ($n_s=0.95$, $r=0.01$ )&& \\
Vilenkin \& Ford (VF-I) & 976.88 & 0.43\\
($n_s=0.95$ ) &&  \\
Vilenkin \& Ford (VF-II) & 978.66 & 0.96\\
($n_s=0.95$ ) & & \\
\\ \hline
\label{thlike}
\footnotetext{Interpretation of $k_c$ depends on the form of model
power spectrum. \\For the recovered spectrum a simple tangent
hyperbolic fit was used.}
\end{tabular}
\end{center}
\end{table}

 Here we have considered a `best fit' cosmology and some finite number
of variations around it~\footnote{ Our analysis here is limited to
flat models cosmological models. We have also ignored the contribution
from tensor perturbations and assumed adiabatic perturbations in this
exploratory paper. Although, inflation generically predicts a
geometrically flat universe, the power spectrum of perturbations in
non-flat universe models has been studied~\cite{nonflatcmb}.  The
effect of tensor perturbations on the CMB anisotropy spectrum is well
studied~\cite{gwcmb} and so is the role of isocurvature
perturbations~\cite{isocurvcmb}. Hence, it is straightforward to
remove these limitations in a more comprehensive future
analysis.}. However, with large but reasonable computational resource,
it is possible to explore a multi-dimensional space of cosmological
parameters to compute a likelihood at each point for an optimal
(recovered) primordial spectrum.

In this work, we have used only the angular power spectrum of WMAP
measured temperature fluctuations (TT). Once the E-polarization power
spectrum (EE) is announced by the WMAP team, our method can be
extended to include TE and EE angular power spectra in obtaining the
primordial power spectrum.  It is also possible extend our recovery of
the primordial spectrum to larger wave-numbers by using the matter
density power spectrum measured by large scale redshift surveys such
as the Sloan Digital sky survey SDSS and $2$degree Field survey,
measurement from Ly-$\alpha$ absorption, and possibly, weak
gravitational lensing in the near future~\cite{max_zal02,gaw99}.

In the absence of a definitive and precise scenario for the generation
of primordial perturbations, a direct inversion of the primordial
spectrum is an extremely appealing approach made possible by the
remarkable quality of the recent cosmological data, in particular, the
anisotropy in the cosmic microwave background temperature.

\acknowledgments We greatly benefited from the exploratory attempt to
solve the problem using a non-iterative method by V. Gopisankararao as
part of the visiting student programme at IUCAA in the summer of
2001. We acknowledge very useful discussions with S. Sridhar and
N. Sambhus regarding the RL deconvolution method.  AS thanks IUCAA for
use of its facilities during his Master thesis work.

\appendix

\section{Some aspects of the method.}

In this appendix, we provide support and justification for some steps
in our method for recovering the primordial power spectrum.  In
section~\ref{initguess} we demonstrate the robustness of the iterative
Richardson-Lucy deconvolution to changes in the initial guess. The
next section (\ref{irl}) discusses the advantage of the improved
Richardson-Lucy method used in our work. In the last section
(\ref{refspec}), we model the broad features of reference spectrum
analytically and show that it is well understood.

\subsection{Effect of the initial guess in the Richardson-Lucy method}
\label{initguess}

\begin{figure}[h]
\includegraphics[scale=0.45, angle=-90]{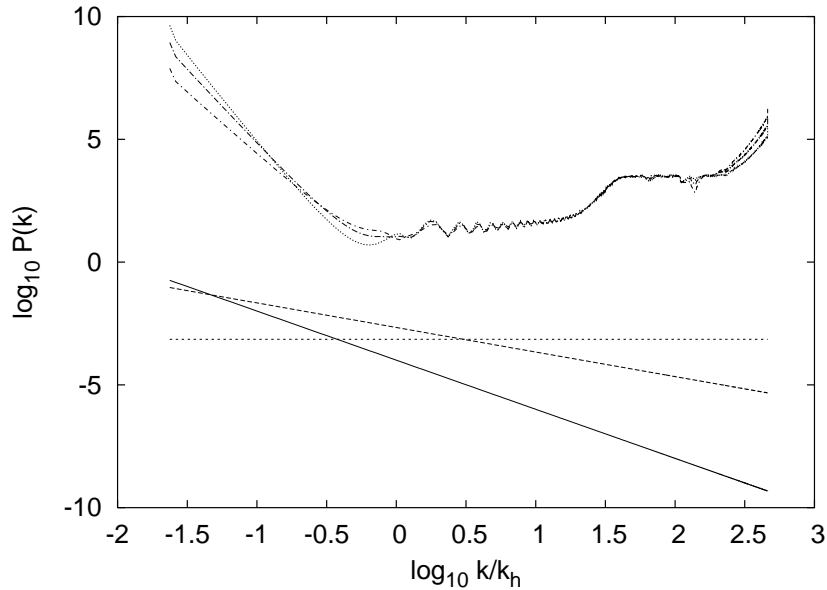}
\caption{ The figure shows the raw power spectrum recovered using the
Richardson-Lucy algorithm starting from three different initial
guesses. The effect of the initial guess is negligible in the other
region $k$ space. As shown in the next section~\ref{refspec}, the
artefacts at low $k$ and high $k$ which are removed by the reference
spectrum have a known dependence on the initial guess. }
\label{guess1}
\end{figure}

The effect of the initial guess is negligible in the Richardson-Lucy
method of deconvolution for our problem. We find that the result for
various different initial guesses come very close to each other after
a few iterations.  Here we demonstrate the robustness of our method by
deconvolving using the $C_l$ from a test spectrum (with a step) using
different initial guesses. In Fig.~\ref{guess1} deconvolved raw $P(k)$
arising from different forms of the initial guess spectra

\be P(k)=constant, \quad P(k)=1/k, \quad P(k)=1/k^2, \ee 

are shown. The effect of the initial guess is absent in the portion of
the $k$ space probed well by the kernel. The deconvolved spectra are
almost identical for all these different initial guesses in that
region of $k$ space. As discussed in section~\ref{refspec}, the
dependence of deconvolved spectrum at the small and large $k$ ends is
well understood and get largely removed when divided by the reference
spectrum.  In a previous work we have checked a large variety of
initial guesses and concluded the RL method applied to our problem of
recovering the primordial spectrum from the CMB anisotropy is
independent of the initial guess.~\cite{arman_msc}.

\subsection{The Improved Richardson Lucy method}
\label{irl}

\begin{figure*} 
\centering
\begin{center} 
\vspace{-0.05in}
\centerline{\mbox{\hspace{0.in} \hspace{2.1in}  \hspace{2.1in} }}
$\begin{array}{@{\hspace{-0.4in}}c@{\hspace{0.3in}}c@{\hspace{0.3in}}c}
\multicolumn{1}{l}{\mbox{}} &
\multicolumn{1}{l}{\mbox{}} &
\multicolumn{1}{l}{\mbox{}} \\ [-0.5cm]
 
\includegraphics[scale=0.32, angle=-90]{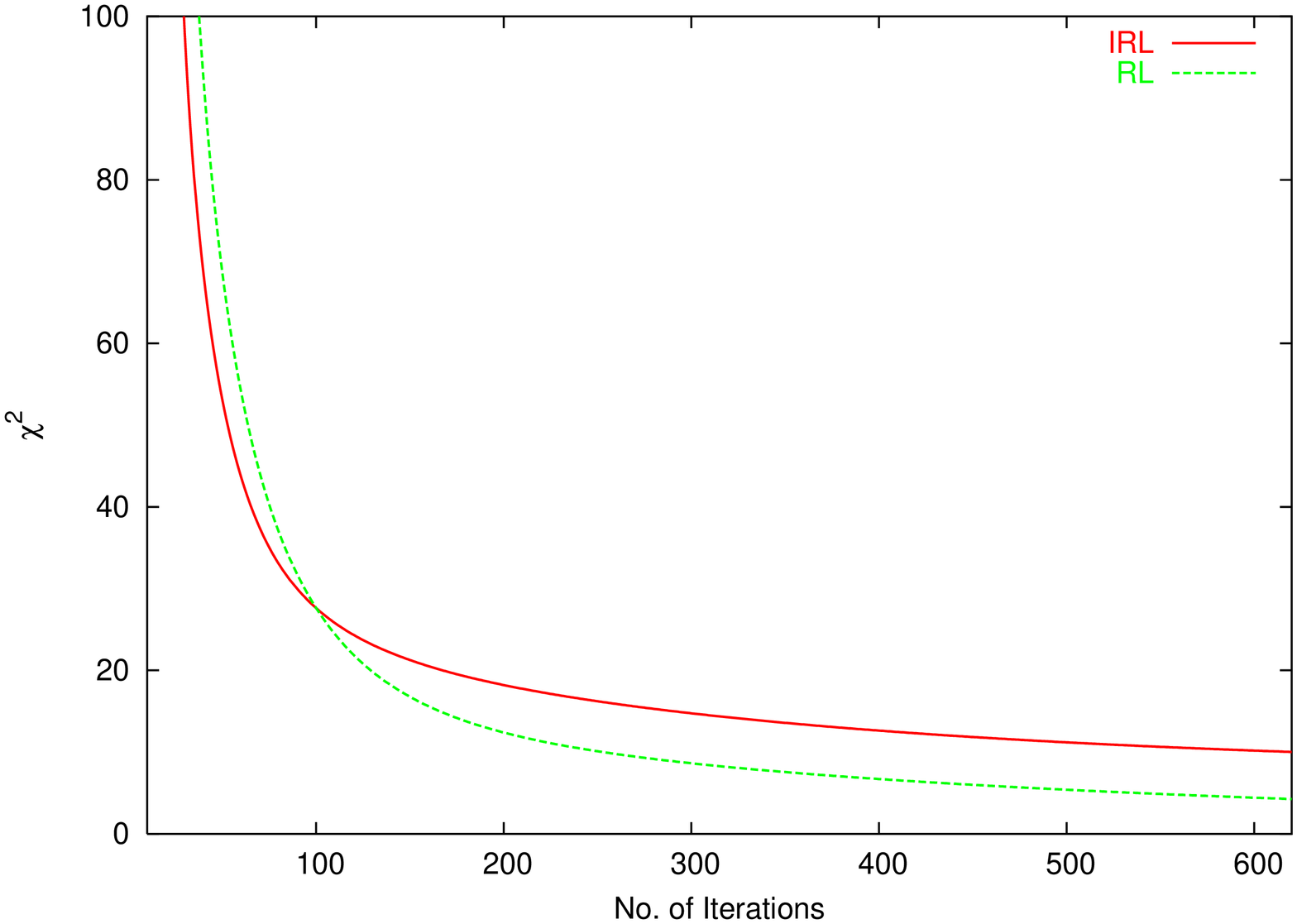}
 
\includegraphics[scale=0.32, angle=-90]{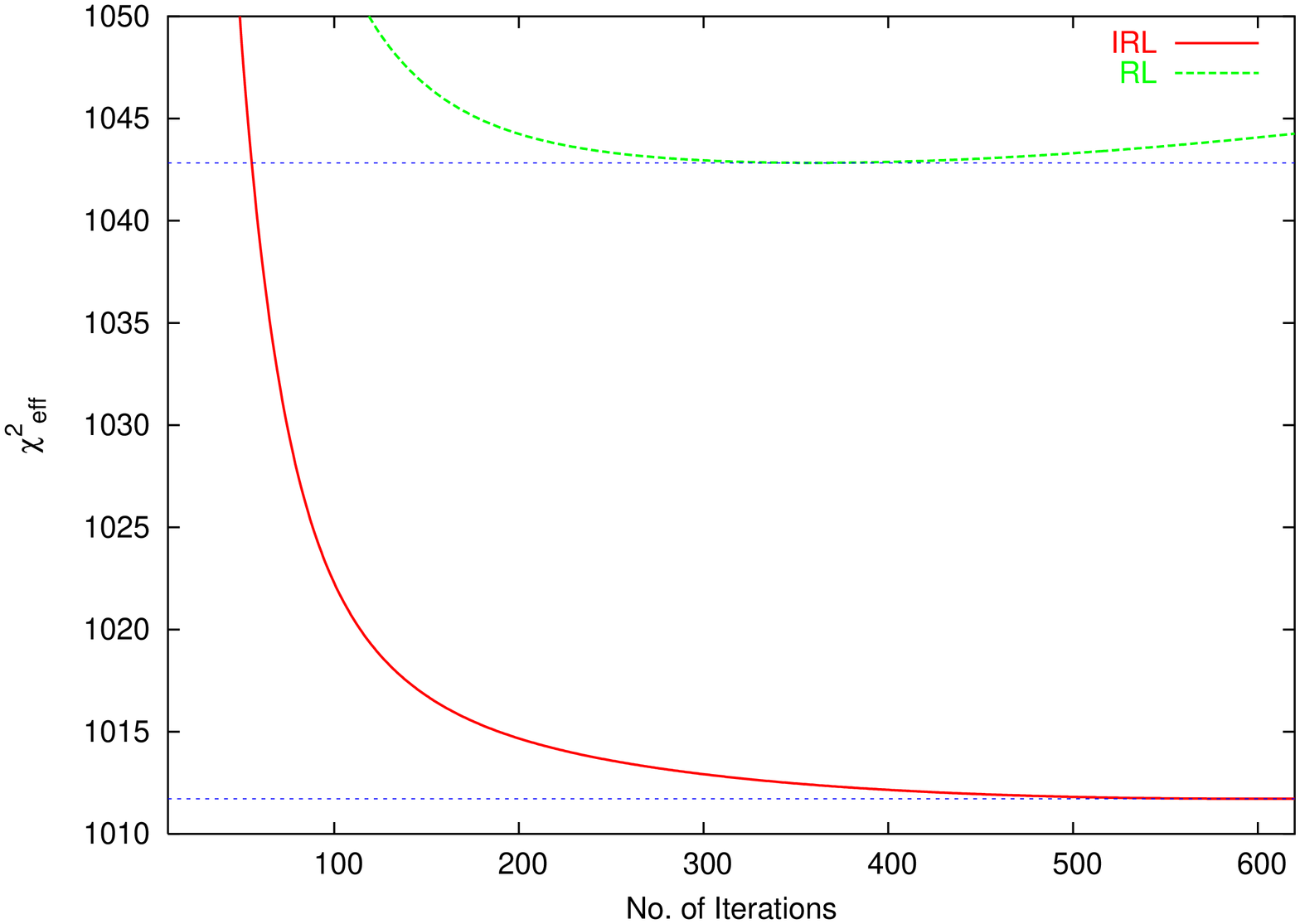}
\mbox{\bf (a)}
\end{array}$

\end{center}
\caption{\small The left panel plots the variation of $\chi^2$ of
$C_l^{(i)}$ obtained after $i$th iteration (w.r.t the binned WMAP
spectrum, $C_l^D$) with increasing iterations for the
Richardson-Lucy (RL) method and improved version (IRL) we present in
this work. The panel on the right, plots the variation of $\chi^2_{\rm
eff}\equiv -2 \ln {\cal L}$ of the $C_l^{(i)}$ given by the WMAP
likelihood, ${\cal L}$.  In contrast to RL, in the IRL method
$\chi^2_{\rm eff}$ converges with iteration and to a significantly
lower value. }
\label{chisqiter}
\end{figure*}

In this section we show the advantage of the improved Richardson Lucy
algorithm given by eq.~(\ref{RLerr}) over the standard method given by
eq.~(\ref{RLstd}). We compare the raw power spectrum $P^{(i)}(k)$ and
the corresponding angular spectrum $C_l^{(i)}$ obtained from the WMAP
binned angular power spectrum, $C_l^D$ using RL and IRL algorithm
as the iterations progress.  In Fig.~\ref{chisqiter}, the left panel
shows the $\chi^2$ of $C_l^{(i)}$ w.r.t $C_l^D$. As expected, the
RL method leads to a lower (better) $\chi^2$ than the IRL method.
However, we are interested in a recovering a $P(k)$ which give $C_l$
that has better likelihood with respect to WMAP data.  The right panel
of Fig.~\ref{chisqiter} plots the WMAP likelihood given in terms of
$\chi^2_{\rm eff}= - 2 \ln {\cal L}$. This figure justifies the
`improved' label to the IRL method. For the RL case, the value of
$\chi^2_{\rm eff}$ is seen to bounce and increase at large
iterations. (This bounce is more pronounced and happens at a lower
iteration for certain cosmological model, e.g., one with high optical
depth to reionization, $\tau$.)  This is a reflection of the problem
of the RL method fitting the noise in the data. In contrast, in the
IRL case the $\chi^2_{\rm eff}$ converges with iterations and to
significantly better (lower) value that for the RL case.

\subsection{Reference spectrum}
\label{refspec}

In this subsection, we show that the reference spectrum used to remove
numerical artifacts from the raw power spectrum recovered by the
deconvolution is analytically well understood. The reference spectrum
reflects the sensitivity of the kernel $|\bar \Delta(l,k)|^2$ to the
$k$ space. (Here, the over-bar in $|\bar \Delta(l,k)|$ alludes to the
binned version related to $\bar G(l,k)$ following
eq.~(\ref{clsum})). In the regions of $k$ space where the kernel
probes $P(k)$ weakly, there is scope for changing the power spectrum
without changing the $C_l$. In such degenerate regions, the Richardson
Lucy method (both RL and IRL) pushes the power spectrum up to the
extent possible changing the $C_l$.

\begin{figure}[h]
\includegraphics[scale=0.45, angle=-90]{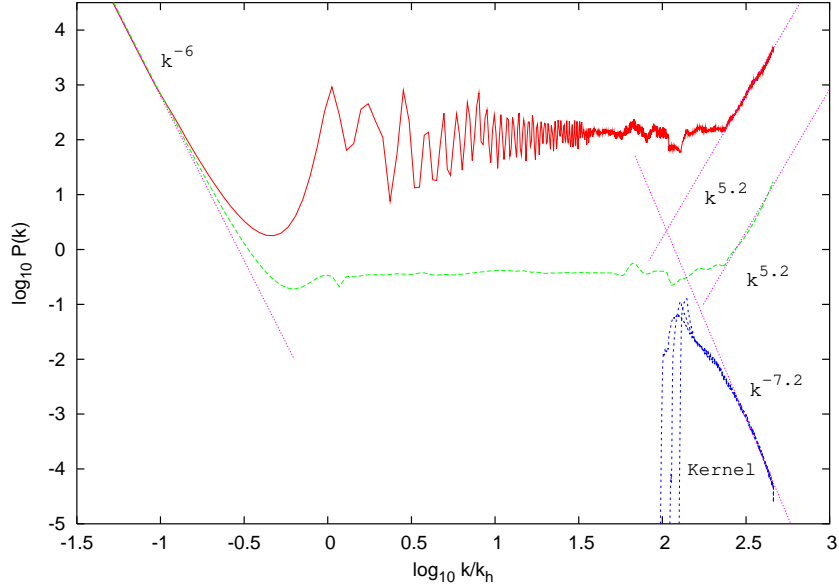}
\caption{The figure shows the $P_{\rm raw}(k)$ and $P_{\rm ref}(k)$
obtained from the WMAP binned data $C_l^D$ for an initial guess
$P^{(0)}(k)\propto 1/k^2$. The dashed straight line labeled corresponds
to the analytical power law form ($k^{-4} P^{(0)}(k) = k^{-6} $) that
matches the identical fall in $P_{\rm raw}(k)$ and $P_{\rm ref}(k)$ at
low $k$. The dashed line matching the rise in $P_{\rm raw}(k)$ and
$P_{\rm ref}(k)$ at large $k$ corresponds to the analytical power law
form ($k^{7.2} P^{(0)}(k)= k^{5.2}$) expected from the roughly $k^{-7.2}$
tail of the kernels for the last few multipole bins.}
\label{refspec_anal}
\end{figure}

Fig.~\ref{refspec_anal} shows that the strong features at the low $k$
and high $k$ are well fitted by analytically obtained power law
forms. At the low wavenumbers ($k/k_h \ll 1$), the CMB anisotropy is
dominated by the Sachs-Wolfe effect.  The sensitivity of the CMB
anisotropy to the power at very small wavenumbers is dominated by the
lowest multipole as $|\bar \Delta(l,k)|^2\propto k^{2 l}$. This is the
well known Grishchuk-Zeldovich effect~\cite{gris_zel78}.  When the
quadrupole is included, $|\bar \Delta(l,k)|^2\propto k^{4}$ is the
strongest probe of the $P(k)$ at low wavenumbers. Hence, there is no
change to $C_l$ (here, chiefly the quadrupole) caused from this region
of $k$ space from the initial guess $P^{(0)}(k)$ onward if
$P(k)\propto k^{-4+\epsilon} \,P^{(0)}(k)$ for $\epsilon >0$. In our
work we have used an initial guess of the form $P^{(0)}(k)\propto
1/k^2$. Hence, the slope of $P(k)$ at low $k$ is driven to the $\sim
k^{-6}$ form shown in Fig.~\ref{refspec_anal}.

At the large $k$ end, the $|\bar \Delta(l,k)|^2$ is governed by the
power slope of the tail in the sensitivity of the highest few
multipole bins shown plotted in Fig.~\ref{refspec_anal}. The slope of
these tails are well fitted by a power law form $k^{-7.2}$. Using the
same argument, it is clear that there will be no change to the $C_l$
caused from this region of $k$ space from the initial guess onward if
$P(k)\propto k^{7.2-\epsilon}\,P^{(0)}(k)$. For the initial guess
$P^{(0)}(k)\propto 1/k^2$, the rise at high $k$ end is driven to a
$\sim k^{5.2}$ form.

The division of reference spectrum used here is a simple numerical
recipe for estimating and removing the artefacts of the deconvolution
method for a given kernel. We have shown that the gross features of
the $P_{\rm ref}(k)$ can be understood analytically. The finer
features of $P_{\rm ref}(k)$ are likely linked to the details of the
structure of $|\bar \Delta(l,k)|^2$ and will be explored in future
work.

%% The reason is apparent from the plot of $G(l,k)$ versus $k$ shown in
%% Fig~\ref{glk}.  At the low $k$ end, the strongest probe is the
%% quadrupole component of the kernel $G(2,k)\sim k^4$ (low $k$
%% asymptotes for the multipoles in Sachs-Wolfe regime scale as
%% $k^{2l}$).  This leads to the $\sim k^4\times P^{(0)}(k)$ factor rise
%% in raw spectrum at low $k$. The sensitivity of the CMB anisotropy to
%% the power at very small wavenumbers ($k/k_h\ll 1$) is proportional to
%% $k^4$ is the well known Grishchuk-Zeldovich
%% effect~\cite{gris_zel78}. At the high $k$ end, the $G(l,k)$ for the
%% highest value of $l_{\rm max}$ used in the deconvolution falls off at
%% large $k/k_h \gsim l$. The rising tail in the raw spectrum at large
%% $k$ reflects this fall off.

\end{document}